%% file: paper.tex
\documentclass[sigconf]{acmart}
\AtBeginDocument{%
  }

\copyrightyear{2025}
\acmYear{2025}
\setcopyright{cc}
\setcctype{by}
\acmConference[CCS '25]{Proceedings of the 2025 ACM SIGSAC Conference on Computer and Communications Security}{October 13--17, 2025}{Taipei, Taiwan}
\acmBooktitle{Proceedings of the 2025 ACM SIGSAC Conference on Computer and Communications Security (CCS '25), October 13--17, 2025, Taipei, Taiwan}\acmDOI{10.1145/3719027.3765145}
\acmISBN{979-8-4007-1525-9/2025/10}

\usepackage{color}
\usepackage{xcolor}
\usepackage{graphicx}
\usepackage[labelformat=simple]{subcaption}
\usepackage{xspace}
\usepackage{multirow}
\usepackage{ulem}
\usepackage{array}
\usepackage{longtable}
\usepackage{balance}
\usepackage{booktabs}    % for \toprule, \midrule, \bottomrule
\usepackage{multirow}    % for \multirow in tables
\usepackage{enumitem}    % to control itemize spacing
\usepackage{lipsum}      % just for dummy text
\usepackage{listings}
\usepackage{algorithm}
\usepackage{algpseudocode}
\usepackage[most]{tcolorbox}

\definecolor{myRed}{HTML}{B85450}
\definecolor{myBlue}{HTML}{2072B8}

\begin{document}

\title{Fingerprinting Deep Packet Inspection Devices \\ by Their Ambiguities}


\author{Diwen Xue}
    \orcid{0000-0002-0616-4765}
    \affiliation{
        \institution{University of Michigan}
        \city{Ann Arbor}
        \state{MI}
        \country{USA}}
    \email{diwenx@umich.edu}

\author{Armin Huremagic}
    \orcid{0009-0001-8976-4044}
    \affiliation{
        \institution{University of Michigan}
        \city{Ann Arbor}
        \state{MI}
        \country{USA}}
    \email{agix@umich.edu}

\author{Wayne Wang}
    \orcid{0000-0000-0000-0000}
    \affiliation{
        \institution{University of Michigan}
        \city{Ann Arbor}
        \state{MI}
        \country{USA}}
    \email{wswang@umich.edu}

\author{Ram Sundara Raman}
    \orcid{0000-0000-0000-0000}
    \affiliation{
        \institution{University of California, Santa Cruz}
        \city{Santa Cruz}
        \state{CA}
        \country{USA}}
    \email{rsundar2@ucsc.edu}

\author{Roya Ensafi}
    \orcid{0000-0003-2188-8267}
    \affiliation{
        \institution{University of Michigan}
        \city{Ann Arbor}
        \state{MI}
        \country{USA}}
    \email{ensafi@umich.edu}

\renewcommand{\shortauthors}{Diwen Xue, Armin Huremagic, Wayne Wang, Ram Sundara Raman, \& Roya Ensafi}

\newcommand{\TODO}[1]{{\hl{#1}}\xspace}
\newcommand{\DX}[1]{{\hl{#1}}\xspace}
\newcommand{\eg}{e.g.\@\xspace}
\newcommand{\ie}{i.e.\@\xspace}
\newcommand{\etal}{et~al.\@\xspace}
\newcommand{\etc}{{\textit{etc.}}\xspace}
\newcommand{\Merit}{\textit{Merit}\xspace}
\newcommand{\TK}{\hl{\bf TK}\xspace}
\newcommand{\tk}{\TK}
\newcommand{\prober}{\textit{Prober}\xspace}
\newcommand{\analyzer}{\textit{Analyzer}\xspace}

\newcommand{\myparagraph}[1]{\smallskip\textbf{#1}\hspace{3pt}}
\newcommand{\nameofthething}{\textit{dMAP}\xspace}

\begin{abstract}

Users around the world face escalating network interference such as censorship, throttling, and interception, largely driven by the commoditization and growing availability of Deep Packet Inspection (DPI) devices. Once reserved for a few well-resourced nation-state actors, the ability to interfere with traffic at scale is now within reach of nearly any network operator. Despite this proliferation, our understanding of DPIs and their deployments on the Internet remains limited---being network intermediary leaves DPI unresponsive to conventional host-based scanning tools, and DPI vendors actively obscuring their products further complicates measurement efforts.

In this work, we present a remote measurement framework, \nameofthething (DPI Mapper), that derives \textit{behavioral fingerprints} for DPIs to differentiate and cluster these otherwise indistinguishable middleboxes at scale, as a first step toward active reconnaissance of DPIs on the Internet. Our key insight is that parsing and interpreting traffic as network intermediaries inherently involves \textit{ambiguities}---from under-specified protocol behaviors to differing RFC interpretations---forcing DPI vendors into independent implementation choices that create measurable variance among DPIs. Based on differential fuzzing, \nameofthething systematically discovers, selects, and deploys specialized probes that translate DPI’s internal parsing behaviors into externally observable fingerprints. Applying \nameofthething to DPI deployments globally, we demonstrate its practical feasibility, showing that even a modest set of 20-40 discriminative probes reliably differentiates a wide range of DPI implementations, including major nation-state censorship infrastructures and commercial DPI products. We discuss how our fingerprinting methodology generalizes beyond censorship to other forms of targeted interference, and we hope our work inspires further measurement efforts toward greater visibility and transparency into DPI devices deployed across the global Internet.

\end{abstract}





\maketitle

\input{01intro}

\input{02background}

\input{03methodology}

\input{04system}

\input{05measurement}

\input{06discussion}
\input{07conclusion}

\begin{acks}

The authors are grateful to the anonymous reviewers for their constructive feedback.
We also thank Brian Huang, Ben Wolin, Ethan McKean, Markian Voronovych for helping with testbed DPI measurements, and Jedidiah Crandall, ValdikSS, and Ivan Nardi for their insightful feedback.
This material is based upon work supported by the National Science Foundation under Grant Numbers CNS-2237552.

\end{acks}

\bibliographystyle{ACM-Reference-Format}
%\balance
\bibliography{refs}

\input{99appendix}

\end{document}

%% file: 01intro.tex
\section{Introduction}
\label{sec:introduction}

The recent decades have witnessed a troubling escalation in network interference faced by users around the world~\cite{accessnowInternetShutdowns}---ranging from outright censorship~\cite{Master2023a, niaki2020iclab, xue2022tspusurvey19} and targeted bandwidth throttling~\cite{xue2021throttlingsurvey18, Anderson2013a} to surreptitious traffic interception or injection of malicious contents~\cite{Raman2020b, marczak2015analysis, citizenlabdpi3}. This trend is fueled in large part by the increasing availability of specialized network equipment, particularly \textit{Deep Packet Inspection} (DPI) devices, which enable monitoring, inspection, and targeted interference with network traffic in real-time. While Internet censorship once limited to a select few well-resourced and motivated nation-state authorities, the commoditization of DPI devices now empowers virtually any ISP or network operator to implement sophisticated filtering and/or interference policies at nearly any network boundary~\cite{filtermap, sundara2020censored}.

\input{figures/dpi}

It is important to acknowledge the dual-use nature of DPI technology: while it serves legitimate functions in network security, its capabilities are also the primary enabler of the network interference we study. Yet, despite their proliferation, our knowledge and visibility into DPI deployments on the Internet remain remarkably limited. For their use as censorship devices, a wealth of literature has measured the \textit{effects} of censorship---documenting which websites are blocked in specific regions, but has done little to illuminate the devices enabling such blocking. This gap arises in part because DPIs operate as network intermediaries rather than endpoints (Figure~\ref{fig:dpi}): they do not often expose open ports on public IPs, nor do they otherwise respond to typical network scanning, making them unsusceptible to standard, host-based scanning tools like Nmap or ZMap~\cite{nmap, durumeric2013zmap}. ``On-path'' DPIs, which act on mirrored traffic out-of-band, may not even be physically inline with routing, so their presence becomes virtually undetectable by standard methods.

Further complicating efforts to measure DPIs is a growing trend among DPI vendors toward obscuring the presence of their devices and avoiding explicit identification. For example, past research often relied on explicit blockpages injected by DPIs---often branded with vendor logos or names---to identify the DPI responsible~\cite{filtermap}. However, following a series of controversies and lawsuits exposing the role of certain Western-made DPI products in facilitating nation-state censorship and surveillance~\cite{citizenlabdpi0, citizenlabdpi2, citizenlabdpi3, dpinews1, dpinews2}, many vendors have now shifted to more generic and indistinct censorship methods, such as silent packet drops or injecting standard TCP Resets. This movement is also encouraged by the Internet's shift towards HTTPS, since DPIs cannot inject blockpages into encrypted traffic. According to public data from Censored Planet~\cite{sundara2020censored} (Figure~\ref{fig:rstblockpage}), one of the largest censorship observatories, censorship by vendor-labeled blockpages has declined by over 85\% over the past six years, replaced by less distinguishable methods. %\textit{How, then, can we perform active reconnaissance on these DPI devices that sit mid-path on the Internet and provide no explicit identifiers?}

In this work, we present a remote measurement framework that derives behavioral fingerprints for DPI devices, enabling researchers to map out, differentiate, and cluster these otherwise indistinguishable middleboxes at scale. The core insight behind this reconnaissance method is that \textit{ambiguities in how network traffic is read provides a source of variance across DPI implementations}. Such ambiguities arise from various factors, such as under-specified corner cases in protocol standards or differing interpretations of RFC guidelines. For example, the IETF specifications do not prescribe a canonical way for reassembling overlapping IP fragments~\cite{rfc6274}, forcing each DPI vendor to independently decide how to handle such scenarios. Our insight is that these implementation-specific decisions, whether explicit or implicit, introduce subtle differences that can be leveraged as behavioral fingerprints for DPIs.

Building on these insights, we design and implement \nameofthething (DPI Mapper), a framework that systematically \textit{discovers}, \textit{selects}, and \textit{deploys} specialized network probes that convert DPIs’ internal traffic parsing ambiguities into externally measurable fingerprints. \nameofthething proceeds in three phases: (1) First, we enumerate a broad range of potentially ambiguous packet sequences with deterministic fuzzing, guided by a comprehensive survey of past DPI evasion literature, to generate a large pool of candidate ``probes'' (\S~\ref{sec:evasionsurvey}). (2) Next, \nameofthething filters and selects the most discriminative probes---those that best differentiate between different DPI implementations---by applying differential analysis against known DPI products (\S~\ref{sec:probeselection}). (3) Finally, \nameofthething conducts large-scale remote probing, observes the behaviors elicited by each probe from DPIs along the network path, and then aggregates these observations across multiple probes into a single behavioral fingerprint for each DPI (\S~\ref{sec:prober} \& \S~\ref{sec:analyzer}).

We apply \nameofthething for large-scale remote fingerprinting of DPI devices across the Internet. We demonstrate the practical feasibility of our fingerprinting approach, finding that even a modest set of 20-40 most discriminative probes yields sufficient variance to reliably distinguish among DPI implementations---even when they employ identical censorship actions (\eg, all injecting indistinct RSTs). We observe that DPI fingerprints \textit{consistently} cluster at the netblock or Autonomous System (AS) level, suggesting that censorship and filtering policies tend to be deployed at these administrative scopes. We identify fingerprint clusters corresponding directly to known nation-state censorship infrastructures (\eg, Iran’s national firewall), as well as globally deployed commercial DPI products such as FortiGate. Perhaps most surprisingly, our results reveal \textit{multiple} fingerprint clusters within some nation-state censorship infrastructures, challenging the longstanding view of them as singular, homogeneous systems. Lastly, tracking fingerprints longitudinally with open-source DPIs (\eg, Suricata, Zeek), we show that their fingerprints remain remarkably stable over multiple years and major releases, highlighting the long-term utility of these behavioral fingerprints.

Our work represents a meaningful step toward greater visibility into the network intermediaries that interfere with users’ traffic on the Internet. While questions and challenges remain in fully understanding these devices, we demonstrate a practical methodology that allows such middleboxes to be remotely measured and differentiated at scale based on their behavioral fingerprints. Importantly, we anticipate our fingerprinting methodology to be \textit{sustainable}---we discuss why the underlying protocol ambiguities enabling it are unlikely to vanish or be easily removed by DPI vendors---and \textit{generalizable}---extending beyond censorship devices to those responsible for other forms of targeted interference such as throttling or MITM attacks. We hope this work inspires further Internet measurement initiatives, collectively advancing the community’s knowledge, transparency, and accountability surrounding these traffic-interfering middleboxes deployed across the global Internet.

%% file: figures/dpi.tex
\begin{figure}[t]
\centering
 \includegraphics[width=\columnwidth,keepaspectratio]{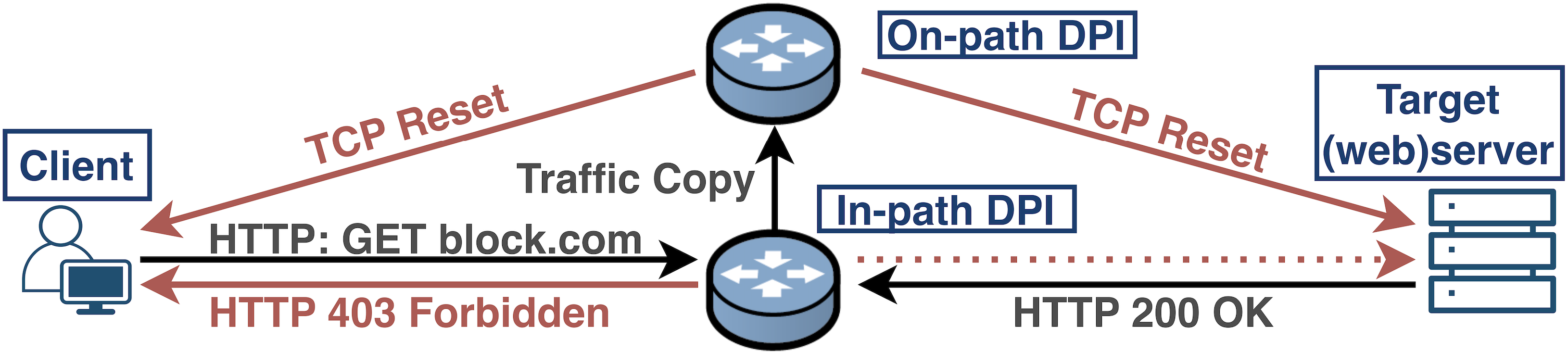}
%\vspace{-15pt}
\caption{DPIs either directly intercept and modify/drop traffic \textit{in-path} or passively monitor mirrored traffic \textit{on-path} and inject packets. \textit{Target} refers to remote endpoint toward which we send probes, with the DPI of interest interfering en route.}
\label{fig:dpi}
%\vspace{-15pt}
\end{figure}

%% file: 02background.tex
\section{Background \& Related Work}
\label{sec:background}

\subsection{Internet Censorship and Interference}

News, anecdotes, and measurement studies collectively suggest that users' Internet traffic is increasingly subject to interference by middleboxes deployed along network paths~\cite{accessnowInternetShutdowns, Master2023a, niaki2020iclab, sundara2020censored}. Among the most prevalent forms of such interference is Internet censorship, which is increasingly practiced by authorities around the world. For over two decades, researchers have conducted both country-specific case studies~\cite{Katira2023a, xue2022tspusurvey19, Padmanabhan2021a, Ramesh2023a} and global-scale censorship measurements~\cite{niaki2020iclab, sundara2020censored, Elmenhorst2021a}, documenting censorship methods ranging from simple IP-based blocking~\cite{Pearce2017a, torprojectPartiallyBlocked}, website filtering~\cite{chinesecheese, globalDNS, howgreatDNS}, and targeted blocking of protocols and circumvention tools~\cite{probetor, probeshadowsocks, GFWfullyencrypted}. Among these, website blocking over HTTP and HTTPS remains the most prevalent and most studied censorship form, typically enabled by DPI devices that inspect the Host header in HTTP requests or the Server Name Indication (SNI) in TLS Clienthellos~\cite{bock2021evensurvey13, vandersloot2018quack, filtermap}.

Beyond censorship, prior work also documented other forms of interference enabled by DPIs, including targeted bandwidth throttling~\cite{xue2021throttlingsurvey18, Anderson2013a}, TLS machine-in-the-middle attacks~\cite{Raman2020b}, malicious traffic injection~\cite{marczak2015analysis} or redirection to malware~\cite{citizenlabdpi3}. This body of research has exposed otherwise covert practices of network interferences and advanced our understanding of adversarial middleboxes in the networks. Yet, relatively few efforts have focused on characterizing and identifying the DPI devices that enable such interference. These DPI devices are the primary focus of our study.

\input{figures/rst_blockpage}

\subsection{Censorship Devices}
Censorship middleboxes can be broadly classified into \textit{in-path} and \textit{on-path} devices~\cite{marczak2015analysis, niaki2020iclab}, as shown in Figure~\ref{fig:dpi}. \textit{In-path} devices operate directly on the network path and can inject, modify, or drop packets en route. \textit{On-path} devices, by contrast, observe a copy of traffic and can inject packets but cannot directly drop or modify them. Our fingerprinting methodology accommodates both device types. These censorship devices conceptually operate using a two-step process~\cite{Xue2025Timing, wails2024precisely}. First, the device \textit{reads} and \textit{interprets} network traffic, extracting information of each flow (\eg, domain names) and evaluating it against preconfigured firewall policies. Next, if a flow is deemed noncompliant with the policy, the device \textit{enforces} censorship through active actions such as dropping packets or injecting TCP resets. As we detail in the next section, our approach leverages implementation-specific differences in the parsing stage (step 1), while relying on the externally observable censorship actions (step 2) to detect when these differences arise.

To date, research measuring censorship devices has largely been approached on a case-by-case, ad-hoc basis. Xue~\etal identified Russia's TSPU by focusing on domain blocklists, under the assumption that devices administered by the same authority would share the same censorship policies~\cite{xue2022tspusurvey19, xue2021throttlingsurvey18}. This approach, however, effectively fingerprints \textit{configurations} rather than \textit{implementations}, since the same device elsewhere could easily load a different set of policies. Dalek~\etal and Marczak~\etal manually engineered network-level features, such as IPIDs, from injected packets to fingerprint Netsweeper and Sandvine DPIs~\cite{citizenlabdpi0, citizenlabdpi1, citizenlabdpi3}. However, such methods are labor intensive and not easily generalizable.
%, and are not guaranteed to be sustainable in the long term (can be removed by vendors with software updates). 
When DPIs expose external-facing IPs, Dalek~\etal also identified them via certain keywords in their banners, but acknowledged these cases might be operator misconfigurations rather than standard practices~\cite{citizenlabdpi1}. In 2020, Raman~\etal proposed clustering DPIs by the blockpages they inject~\cite{filtermap}. However, this method relies on the DPI actively injecting user-visible blockpages---an increasingly rare behavior, given many devices now favor less overt censorship actions (\eg, generic RST injection), as shown in Figure~\ref{fig:rstblockpage}.

%The work most related to ours are Autosonda and Cenfuzz~\cite{jermyn2017autosondasurvey31, raman2022cenfuzzsurvey7}, both of which perform mutations to HTTP or TLS requests to study the rules and triggers of censorship devices. Yet, both focused on specific geographic regions and primarily examined application-layer features. We extend these work by developing a scalable, generalizable approach spanning the network stack for systematically measuring and fingerprinting DPIs globally. 

The work most related to ours are Autosonda and Cenfuzz~\cite{jermyn2017autosondasurvey31, raman2022cenfuzzsurvey7}, both of which perform mutations to HTTP or TLS requests to study the rules and triggers of censorship devices. Our primary novelty compared to them lies in a broader and more systematic exploration of fingerprintable ambiguities across network layers. Both~\cite{jermyn2017autosondasurvey31, raman2022cenfuzzsurvey7} exclusively targeted application-layer features (HTTP), while \nameofthething generalizes across network/transport/application layers, where parsing behavior is less likely to be affected by site-specific configuration policies. Furthermore, \cite{raman2022cenfuzzsurvey7} focused on only one geographic region, and~\cite{jermyn2017autosondasurvey31} performed local measurements within a single city. We extend these work by developing a scalable, generalizable approach spanning the network stack for systematically measuring and fingerprinting DPIs globally.

%% file: figures/rst_blockpage.tex
\begin{figure}[t]
\centering
 \includegraphics[width=\columnwidth,keepaspectratio]{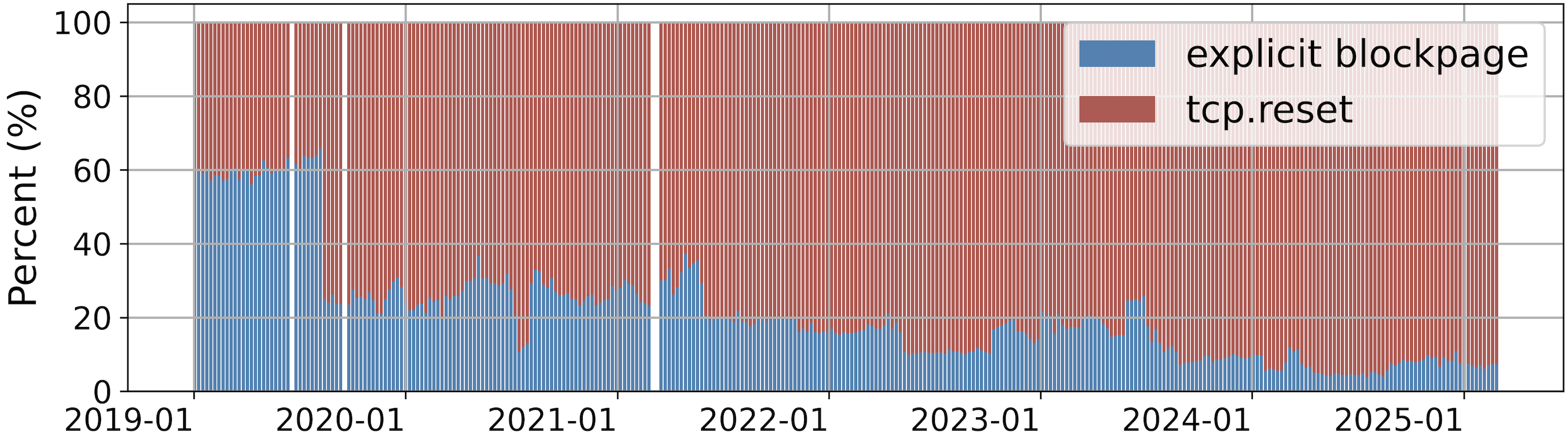}
%\vspace{-15pt}
\caption{Different censorship actions observed in Censored Planet's HTTP measurements~\cite{sundara2020censored}. Explicit blockpages have increasingly been replaced by less overt RST injections.}
\label{fig:rstblockpage}
%\vspace{-15pt}
\end{figure}

%% file: 03methodology.tex
\section{Methodology}
\label{sec:methodology}

\input{figures/dpi_workflow}

The goal of this work is to develop a methodology for fingerprinting DPI devices in a way that allows researchers to differentiate different DPI implementations and cluster similar ones. Our methodology is designed under the following constraints and objectives:

\textbf{Generality}: The technique must be universally applicable to \textit{any} DPI that filters traffic at the network level. Importantly, we do not make assumptions regarding the mechanisms these devices use to enforce censorship---whether through explicit RST injections, blackholing, or other methods. We assume the DPI may not produce self-identifying artifacts (\eg, a blockpage with vendor name).

\textbf{Black-Box Assumption}: We treat DPIs as black boxes, assuming no prior knowledge of their internal logic, configuration, or codebase. Remote fingerprinting must work \textit{without} physical access to the DPI or its internal states, but rely solely on externally observable feedback (\ie, pass/block of the connection).

\textbf{Fingerprinting Implementation, Not Deployments}: The characteristics leveraged for fingerprinting should reflect inherent aspects of the DPI's implementation---such as how it parses packets, reassembles fragments, or manages TCP states---rather than its site-specific configuration or policies (\eg, which domains are blocked). While the latter may differ significantly between deployments of the same DPI product, implementation-specific behaviors tend to remain consistent across deployments.

In \S~\ref{sec:ambiguities}, we describe the high-level intuition behind exploiting ambiguities in packet parsing and flow reassembly to elicit \textit{implementation-specific} behavior from different DPIs. Then, in \S\ref{sec:evasion}, we discuss the relationship between these ambiguities and classic DPI evasion attacks, showing how known evasion strategies often hinge on the very same parsing discrepancies that provide a source of variance for fingerprinting. Finally, we survey prior DPI-evasion literature and catalog the ambiguities exploited, which help inform the discovery and selection process of our fingerprinting probes.
%Building on these insights, we describe \nameofthething in the next section.

%Building on that insight, \S\ref{sec:differential-fuzzing} describes how we systematically translate classes of known ambiguities learned from prior work into a \emph{deterministic differential fuzzing} framework to generate candidate probes that likely lead to differentiating responses across DPI devices.

\subsection{Overview: Fingerprinting by Ambiguities}
\label{sec:ambiguities}

Despite their varying implementations, the high-level intended functionality of DPIs follows a broadly similar workflow, as shown in Figure~\ref{fig:dpiworkflow}: (1) Traffic Ingestion---capture inline or mirrored traffic; (2) Parsing, Processing, and Content Inspection---analyze packet structures, optionally perform flow reassembly and track connection states; (3) Policy Evaluation---determine whether traffic should be blocked according to predefined rules; and (4) Enforcement---actively implement blocking measures when necessary.

Previous work on fingerprinting DPIs often leveraged either policy information (\eg, domain blocklists) or explicit artifacts from the enforcement step (\eg, identifiable blockpages~\cite{filtermap}, HTTP headers containing vendor information~\cite{citizenlabdpi1, citizenlabdpi2}, or signatures of the injected packets like fixed IPIDs~\cite{citizenlabdpi3}). While these artifacts provide straightforward fingerprints, they lack generalizability, as they depend heavily on injections with identifying information that many DPIs increasingly avoid, opting instead for more indistinct methods like simple packet dropping or generic TCP RSTs.

\textit{At the core of our approach is to shift the focus from how DPIs \textbf{\emph{write}} to how they \textbf{\emph{read}}}. Instead of relying on detectable signatures from packet injections, we fingerprint DPIs based on how they parse, inspect, and \textit{interpret} network traffic. This approach applies universally, as all DPIs by definition must inspect traffic to evaluate policies, regardless of their specific method of enforcement.

For any fingerprinting method to be effective, the behavior it measures must exhibit enough variance across different implementations. Although all DPIs ideally follow the same conceptual process of reading traffic, in practice, traffic parsing and interpretation contain various ambiguities that can lead to \textit{inconsistent handling of the same packet sequence among different DPI products}~\cite{handley2001networksurvey1, ptacek1998insertionsurvey2}. These ambiguities generally arise from three factors: first, many DPIs implement only a subset of the full protocol specifications, often optimizing for throughput or simplicity~\cite{handley2001networksurvey1}. For example, while many modern DPI engines track TCP connection states, they may not handle edge cases such as ``Simultaneous Open'', where both client and server send SYN packets during the handshake. Previous work has shown that some DPIs fail to properly initialize states when encountering these non-standard handshakes~\cite{xue2022tspusurvey19}. As another example, some DPIs take shortcuts when parsing packets, such as assuming certain packet header lengths rather than dynamically parsing them~\cite{tcptimestampgfwsurvey29}, leading to divergent behaviors if the header size deviates from those assumptions.

Second, protocol RFCs often leave certain operational details such as resource management under-specified, forcing implementations to make their own design decisions. For example, RFC791 defines IP fragmentation reassembly behaviors but does not specify how to manage buffers and buffer sizes during the process. Different DPI products can adopt different buffer size limits, and these limits in turn provide variability that helps differentiate one implementation from another. Figure~\ref{fig:fragsize} shows a motivating example.

Finally, even when RFCs aim to specify behavior, the use of natural language inherently leaves room for varied interpretations, especially under rare edge cases. Prior studies using natural language processing have identified many potential ambiguities in various network protocols~\cite{rfcnlpambiguities}. In \S~\ref{sec:rootcause}, we found that certain leveraged fingerprints were indeed caused by developers' differing interpretations. Collectively, these ambiguities can lead to divergent implementation choices across different vendors, providing high-variance, measurable differences that form the core of our fingerprinting approach.

\subsection{Evasion-By-Ambiguities}
\label{sec:evasion}

\textit{How do we find these ambiguities?} In principle, one could design ambiguity-based probes for DPI fingerprinting through multiple approaches, ranging from exhaustively analyzing RFC languages to brute-force fuzzing of every possible packet-field combination. In this work, however, we choose to ground our probe design in classic DPI evasion attacks. Two considerations motivate this approach:

\textit{Higher Likelihood of Divergence:} First, evasion attacks highlight precisely those ambiguities in the protocol where real DPI implementations diverge from ideal or reference behaviors. An evasion attempt succeeds when one DPI accepts and reassembles an ambiguous packet sequence that an end host mis-parses or drops; \ie, these attacks exist precisely because implementations diverge under those conditions. Additionally, empirical evidence has shown that at least 86.74\% (and up to 100\%) of evasion strategies effective against one DPI fail against another~\cite{moon2024prydesurvey20}. 

\input{figures/fragsize}

\textit{Observable Feedback in a Black-Box Setting:} In remote DPI fingerprinting, we treat DPIs as blackboxes and assume no access to its internal states. That means the only feedback we can reliably measure is a binary signal---whether the connection continues or has been blocked. Evasion attacks, by definition, exploit exactly those ambiguities that produce such observable differences when interpreted differently. For example, consider the ambiguity around fragmentation buffer size. An evasion attempt (Figure~\ref{fig:fragsize}) might split a sensitive request into $N$ fragments, succeeding only against DPIs with reassembly buffers smaller than $N$ but failing to evade a different DPI with a larger buffer. Observing the differing blocking outcomes indirectly reveals the underlying implementation-specific buffer limit, which can be used for fingerprinting. 

Therefore, our approach first surveys existing DPI-evasion literature to identify ambiguities already validated in real-world censorship contexts for causing \textit{observable} and \textit{divergent} DPI behaviors.

\input{tables/evasionsurvey}

\subsection{A Survey of DPI Evasion Attacks}
\label{sec:evasionsurvey}

To systematically identify promising ambiguities for DPI fingerprinting, we surveyed 31 prior works on DPI evasion attacks targeting open-source, commercial, and nation-state DPI systems. Inspired by the seminal work of Handley \etal and Ptacek \etal~\cite{handley2001networksurvey1, ptacek1998insertionsurvey2}, we adopt a simple taxonomy that catalogs evasions according to the network layer where the ambiguity arises (IP, TCP, or application) and whether the ambiguity is intra-packet (related to the parsing of fields) or inter-packet (connection tracking, reassembly, \etc). It's important to note that not all DPI evasions are due to ambiguous traffic interpretations. For example, we exclude evasions involving TTL-limited packets, as TTL expiration follows a well-defined protocol behavior that offers little scope for divergent interpretations. Similarly, we exclude evasions based on orthogonal mechanisms such as encrypted tunneling, which bypasses rather than exploits parser discrepancies. Additionally, in line with our measurement scope in \S~\ref{sec:measurement}, we restrict application-layer evasions to HTTP(S), excluding those targeting other protocols like DNS~\cite{nourin2023measuringsurvey10, harrity2022getsurvey11}.

Table~\ref{tab:evasionsurvey} provides an overview of the ambiguities commonly exploited in previous DPI evasion attacks. From a fingerprinting perspective, this table highlights the aspects of packet sequences most likely to expose divergent DPI behaviors. Specifically, we see recurring emphasis on three main categories of ambiguities: (1) the parsing of malformed or partially invalid packet fields; (2) the handling of fragmentation reassembly (at IP, TCP, and TLS layer), where differences in buffer limits or overlap resolution can cause misalignment in how DPIs see stream contents; and (3) the management of TCP states, especially unusual packet sequences that deviate from the typical TCP handshake/teardown procedures. Within each category, the survey also pinpoints specific fields that appear to be more prone to divergent interpretation (\eg, TCP flags, IP options, and sequence numbers, as opposed to IP addresses or ports). In \S~\ref{sec:generateprobes}, we build on these insights to design a deterministic fuzzing framework that generates candidate fingerprinting probes by permuting the most ambiguity-prone aspects of network traffic.

It's worth emphasizing that one dimension we do not consider in our survey is the \textit{effectiveness} of each evasion attack. Although this metric is central to many evasion studies, our goal in this work is \textit{not} to evade DPIs, but to identify ambiguities that introduce measurable variances across implementations. Indeed, a ``perfect'' evasion attack that bypasses a broad range of DPIs reveals little information about \textit{which} DPI a network is actually using. 

Lastly, while we do not claim our survey to be exhaustive, the high degree of overlap among prior works leads us to believe that we have captured the classes of ambiguities that are most relevant for our fingerprinting efforts. Because we focus on the underlying ambiguities rather than reusing evasion attacks verbatim, completeness in cataloging every specific evasion sequence is not essential. For example, no fewer than 16 studies describe attacks tied to TCB teardown ambiguities; even if we missed specific packet sequences proposed in more recent works, those techniques typically rely on the same underlying issues in state management. In the next section, we describe the design of \nameofthething, our framework that implements this methodology for discovering, selecting, and deploying probes to fingerprint DPIs based on their handling of protocol ambiguities.

%% file: figures/dpi_workflow.tex
\begin{figure}[t]
\centering
 \includegraphics[width=\columnwidth,keepaspectratio]{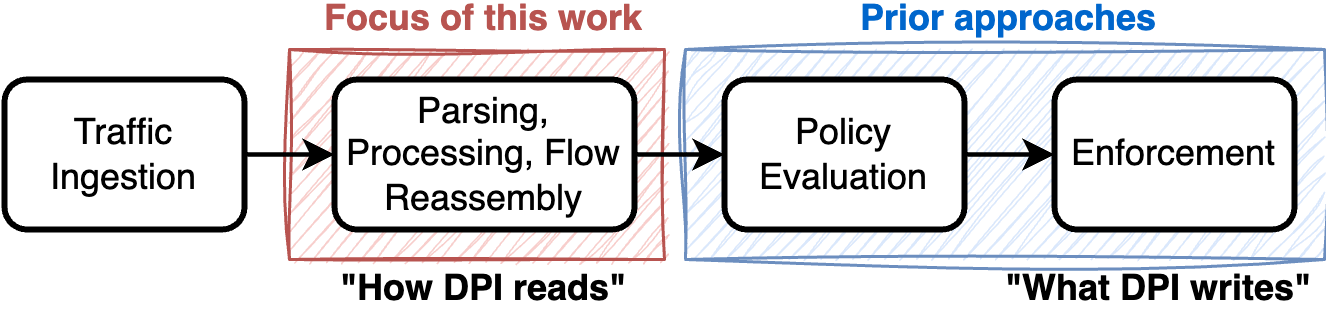}
%\vspace{-15pt}
\caption{Prior fingerprinting efforts focus on installed policies or artifacts from the enforcement step. We instead fingerprint DPIs based on how they parse and interpret traffic.$\diamond$}
\label{fig:dpiworkflow}
%\vspace{-15pt}
\end{figure}

%% file: figures/fragsize.tex
\begin{figure}[t]
\centering
 \includegraphics[width=\columnwidth,keepaspectratio]{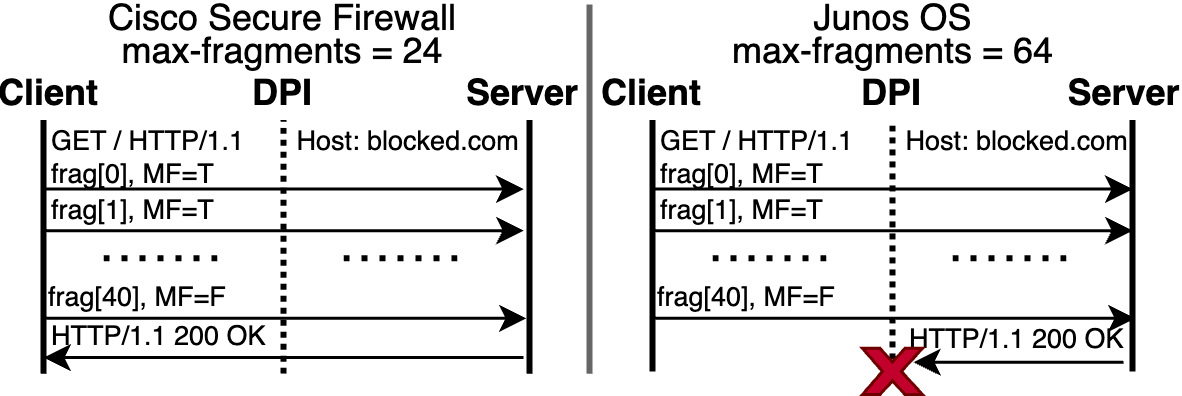}
%\vspace{-15pt}
\caption{A motivating example where a fragmentation-based evasion attempt may succeed or fail based on the DPI's internal reassembly buffer size, making its handling of this ambiguity externally fingerprintable.}
\label{fig:fragsize}
%\vspace{-15pt}
\end{figure}

%% file: tables/evasionsurvey.tex
\begin{table*}[t!]
\footnotesize
\centering

\begin{tabular*}{2\columnwidth}{@{\extracolsep{\fill}}m{0.5cm}m{1.9cm}m{3.0cm}m{6.7cm}m{4.5cm}}

\toprule
        {Layer} & {Type} & {Ambiguity} & {Examples/Specifics} & {Reference} \\
\midrule

IP & Fragmentation & IP Fragmentation Reassembly & - in-order/out-order/overlapping fragments &  \\ 
 &  &  & - max timeout / number of fragments in reassembly buffer &  \\ 
 &  &  & - max disorder allowed (ipfrag\_max\_dist) &  \multirow{1}{*}{\cite{handley2001networksurvey1, ptacek1998insertionsurvey2, khattak2013towardssurvey6, bock2019genevasurvey9, shankar2003activesurvey17, xue2022tspusurvey19}} \\ \midrule

IP & Fragmentation  & IP Fragmentation Semantics & - invalid fragmentation offset; invalid MF/DF flag &  \\ 
 &  &  & - min fragment size acceptable for reassembly &  \cite{handley2001networksurvey1, ptacek1998insertionsurvey2, shankar2003activesurvey17} \\  \midrule

IP & Packet Parsing  & Malformed Header Processing &  - invalid IP options (type/value) &  \\ 
 &  &  &  - invalid proto; reserved bit &  \cite{handley2001networksurvey1, ptacek1998insertionsurvey2, li2017liberatesurvey5} \\  \midrule

TCP & Packet Parsing  & TCP Header Processing & - invalid TCP Flag combination (syn/rst, ack not set, \etc)&  \\ 
 &  &  &  - invalid window / windowscale; urgent pointer processing &  \multirow{1}{*}{ \cite{handley2001networksurvey1, ptacek1998insertionsurvey2, wang2017yourstateisnotminesurvey3, wang2020symtcpsurvey4, li2017liberatesurvey5, khattak2013towardssurvey6, bock2020exposingsurvey12, bock2020comesurvey14, shankar2003activesurvey17, genevaindiasurvey24, amich2023deresistorsurvey25, zhang2022statediversurvey30}} \\ \midrule

TCP & Packet Parsing  & TCP Option Processing & - invalid TCP option type&  \\ 
 &  &  &  - unsolicited MD5; invalid timestamp; fast open processing & \multirow{1}{*}{\cite{handley2001networksurvey1, ptacek1998insertionsurvey2, wang2017yourstateisnotminesurvey3, wang2020symtcpsurvey4, bock2019genevasurvey9, shankar2003activesurvey17, amich2023deresistorsurvey25, tcptimestampgfwsurvey29, zhang2022statediversurvey30}} \\ \midrule

TCP & Conn Tracking  & TCB Creation & - packet sequence creating TCB at DPI (single SYN, single ACK, \etc) & \cite{ptacek1998insertionsurvey2, wang2017yourstateisnotminesurvey3, khattak2013towardssurvey6, gfwtechnicalreportsurvey8, bock2020exposingsurvey12, bock2021evensurvey13, bock2020comesurvey14, xue2022tspusurvey19, moon2024prydesurvey20} \\ \midrule
 
TCP & Conn Tracking & TCB Re/de-synchronization & - packet sequences that re-synchronizes/reverses TCB &  \cite{wang2017yourstateisnotminesurvey3, bock2019genevasurvey9, bock2020exposingsurvey12, bock2020comesurvey14, xue2022tspusurvey19, moon2024prydesurvey20, spoofdpisurvey22} \\ \midrule

 TCP & Conn Tracking & TCB Teardown & - packet sequences that tears down TCB maintained at DPI &  \\ 
 &  &  & - timeouts, or max number of packets examined in a flow &  \multirow{1}{*}{\cite{ptacek1998insertionsurvey2, wang2017yourstateisnotminesurvey3, wang2020symtcpsurvey4, li2017liberatesurvey5, khattak2013towardssurvey6, gfwtechnicalreportsurvey8, bock2019genevasurvey9, bock2020comesurvey14, bock2020detectingsurvey15, xue2021throttlingsurvey18, xue2022tspusurvey19, moon2024prydesurvey20, genevaindiasurvey24, amich2023deresistorsurvey25, ververis2021understandingsurvey26, zhang2022statediversurvey30}} \\ \midrule

TCP & Conn Tracking & TCP Stream Reassembly & - invalid seq/ack number (seq < ISN, premature/duplicate ack, \etc) & \multirow{1}{*}{\cite{handley2001networksurvey1, ptacek1998insertionsurvey2, wang2017yourstateisnotminesurvey3, wang2020symtcpsurvey4, li2017liberatesurvey5, bock2020exposingsurvey12, bock2020detectingsurvey15, shankar2003activesurvey17, moon2024prydesurvey20, goodbyedpisurvey21, amich2023deresistorsurvey25, zhang2022statediversurvey30}} \\ \midrule

TCP & Fragmentation & TCP Segmentation & - overlapping segments (partial/whole, in-order/out-order) &  \\
 &  &  & - min segment size; max number of segments allowed & \multirow{1}{*}{\cite{handley2001networksurvey1, ptacek1998insertionsurvey2, wang2017yourstateisnotminesurvey3, khattak2013towardssurvey6, bock2019genevasurvey9, nourin2023measuringsurvey10, bock2021evensurvey13, bock2020comesurvey14, xue2021throttlingsurvey18, xue2022tspusurvey19, goodbyedpisurvey21, spoofdpisurvey22, genevaindiasurvey24, amich2023deresistorsurvey25}} \\ \midrule

HTTP & Request Parsing & Request Line Parsing & - invalid HTTP version / method &  \\
 &  &  & - additional spaces/tabs; alternative delimiters  &  \\
 &  &  & - multiple requests in TCP packet; keyword location within request & \multirow{1}{*}{\cite{raman2022cenfuzzsurvey7, nourin2023measuringsurvey10, harrity2022getsurvey11, goodbyedpisurvey21, ververis2021understandingsurvey26, yadav2018lightsurvey27, jermyn2017autosondasurvey31}} \\ \midrule

HTTP & Request Parsing & Host Header Parsing & - keyword/hostname permutation (capitalize, remove, pad, alternate) & \multirow{1}{*}{\cite{khattak2013towardssurvey6, nourin2023measuringsurvey10, harrity2022getsurvey11, kakhki2016bingeonsurvey16, goodbyedpisurvey21, yadav2018lightsurvey27, jermyn2017autosondasurvey31}} \\ \midrule

TLS & TLS Record Parsing & TLS Record Semantics & - Prepending CH records with other TLS records & \multirow{1}{*}{\cite{xue2021throttlingsurvey18, xue2022tspusurvey19}} \\ \midrule

TLS & Fragmentation & TLS Record Fragmentation & - fragment CH record into multiple TLS fragments &  \multirow{1}{*}{\cite{niere2023postersurvey23, identifyingsslsurvey28}} \\ \midrule

TLS & Clienthello Parsing & Clienthello Parsing & - CH fields permutation (ciphersuite, version); SNI permutation & \multirow{1}{*}{\cite{raman2022cenfuzzsurvey7}} \\ \midrule

IP/TCP & Malformed packet & Checksum & - invalid IP/TCP checksum & \multirow{1}{*}{\cite{handley2001networksurvey1, ptacek1998insertionsurvey2, wang2017yourstateisnotminesurvey3, li2017liberatesurvey5, bock2019genevasurvey9, bock2020exposingsurvey12, shankar2003activesurvey17, amich2023deresistorsurvey25}} \\ \midrule

IP/TCP & Malformed packet & Length Fields & - invalid length fields in IP/TCP/TLS headers/options & \multirow{1}{*}{\cite{handley2001networksurvey1, ptacek1998insertionsurvey2, wang2017yourstateisnotminesurvey3, li2017liberatesurvey5, bock2019genevasurvey9, shankar2003activesurvey17, amich2023deresistorsurvey25}} \\ \midrule

\bottomrule

\end{tabular*}

\caption{\textbf{A Survey of DPI Evasion Attacks.} Overview of common ambiguities exploited, categorized by network layer and type. $\diamond$}
\label{tab:evasionsurvey}
%\vspace{-15pt}
\end{table*}

%% file: 04system.tex
\section{\nameofthething Architecture and Experimentation}
\label{sec:system}

\input{figures/cendpi}

Figure~\ref{fig:cendpi} presents an overview of \nameofthething, our measurement framework for fingerprinting DPI devices. We begin by describing the probing module in \S~\ref{sec:prober}, which builds packets according to the probe configurations provided and performs parallel measurements against targets. The collected results are then processed by \analyzer (\S~\ref{sec:analyzer}), which interprets the raw responses looking at both the control and test measurements and producing a single verdict for each probe-target pair. We defer our discussion on how specific probes are selected to \S~\ref{sec:probeselection}, where we describe how we fuzz candidate probes based on ambiguities identified earlier, and how we select the most discriminative probes with differential analysis.

\subsection{\prober}
\label{sec:prober}

\prober is the centerpiece of the \nameofthething framework, responsible for crafting and sending network probes to measure DPI behaviors. It takes as input a list of probe configurations, each specifying a single measurement (\ie, a single transport-layer connection). Every probe configuration, written in YAML format, defines a precise sequence of packets to be sent, starting from the TCP handshake. For each packet, the configuration allows fine-grained control over fields spanning all protocol layers, from Ethernet to the application layer, including both header fields and payload contents. Additionally, each packet definition may optionally specify whether \prober should wait for a response before sending the next packet (\eg, waiting for a SYN-ACK to learn the server's ISN). Note that the destination IP and port, as well as the domain name used in the application-layer request, are \textit{not} fixed within the probe configuration itself but are supplied at runtime, which allows a single probe configuration to be reused across multiple measurements. An example of a probe configuration is provided in the Appendix~\ref{sec:exampleprobe}.

Alongside probe configurations, \prober also takes a list of \textit{Targets}. It should be noted that for our measurements, the \textit{Targets} themselves are \textit{not} the DPI we aim to fingerprint; rather, they are normal web servers situated behind the DPI of interest, which sits upstream of the web server and intercepts and filters traffic between \prober and the \textit{Target}. For each \textit{Target}, we specify both a Control Domain (\eg, example.com), expected to pass through the DPI unblocked, and a Test Domain (\eg, blocked.com), which is expected to trigger the DPI's filtering. The list of \textit{Targets} can either be measured in an initial discovery phase or sourced from open datasets released by censorship observatories such as Censored Planet~\cite{sundara2020censored}.

At runtime, \prober parallelizes measurements across \textit{Targets}, but enforces a strict sequential ordering between consecutive probes. To support large-scale experiments, measurements using the same probe configuration but targeting different \textit{Targets} may reuse the same source port, provided that no two transport-layer connections within the same measurement run share the same four-tuple. This prevents ``residual censorship'' (lingering blocking state associated with a particular connection identifier) from affecting subsequent probes. As an additional precaution, we introduce a 120-second delay between consecutive probes (even when their four-tuples are different), a timeout previously found sufficient to clear most residual censorship effects~\cite{vandersloot2018quack}. For each \textit{Probe-Target} pair, \prober first launches a Control Measurement using the Control Domain in the application-layer request. Next, it runs Test Measurement using the same packet sequence but substituting the Test Domain. For each measurement, \prober dynamically tracks the \textit{expected} TCP SEQ/ACK numbers, incrementing them based on both outgoing and incoming traffic. This allows us to specify relative SEQ/ACK numbers in probe configurations without having to know the \textit{Target}'s runtime ISN choice. Finally, for each measurement, \prober logs all packets sent and received (in JSONL or PCAP) for analysis.

\prober represents the most significant engineering effort in \nameofthething, with over 4000 LOC excluding probe configurations. It is designed to be highly flexible, supporting arbitrary packet sequences and fine-grained mutations across protocol layers. Due to space constraints, we omit a full description of \prober's implementation but refer interested readers to our open-source repository (\S~\ref{sec:openscience}).

\subsection{Analyzer}
\label{sec:analyzer}

Each entry in \prober's raw output corresponds to a single measurement, recording the packets exchanged during the measurement. \analyzer begins by parsing these packet traces and performing a minimal sanity check (\eg, discarding any measurement that does not contain a completed TCP handshake). Next, \analyzer annotates each valid measurement by examining the immediate response after sending the application-layer request, such as explicit TCP resets, blackholing (absence of further packets), or, in cases where an application-layer response is present, whether the response body matches any known blockpage signatures\footnote{We try to match the response body against a curated open database of known censorship blockpages maintained by Censored Planet~\cite{sundara2020censored}.}. %Importantly, at this stage, \analyzer only logs these raw observations; it does not yet assign a verdict such as ``blocking'' or ``no blocking''.

Once individual measurements have been annotated, \analyzer groups them by the tuple (\textit{Target, Probe, isControl}), where \textit{isControl} indicates whether the measurement used the Control Domain (expected not to trigger censorship) or the Test Domain (expected to trigger censorship). Typically, we repeat multiple measurements for each tuple to account for transient network variations. Within each group, \analyzer consolidates individual annotations into a single outcome by prioritizing explicit signals over more ambiguous ones, based on their relative reliability as indicators of DPI interference. For example, explicit responses like matched HTTP blockpages are more conclusive than mere RSTs, which in turn take precedence over blackholing. Likewise, any clear evidence of blocking outweighs ``no blocking'', under the assumption that DPIs might occasionally fail to block when they should (\eg, under heavy load), but will rarely produce a conspicuous blockpage by accident.

Next, for each (\textit{Target, Probe}) pair, \analyzer compares four consolidated measurement results to interpret how the mutation introduced by the \textit{Probe} affects the DPI's handling of traffic: \textbf{R1}(Reference Control)---Standard request without any mutation using Control Domain. \textbf{R2}(Reference Test)---Standard request using Test Domain. \textbf{R3}(Mutated Control)---Mutated packet sequence supplied with Control Domain. \textbf{R4}(Mutated Test)---Mutated packet sequence with Test Domain. Table~\ref{tab:analyzer} enumerates common patterns~\footnote{Table~\ref{tab:analyzer} shows 8 out of 15 possible combinations and their interpretations. Of the remaining 7 combinations, 5 involve \{R1,R2\}, indicating failures in Control measurements, thus invalidating further analysis. The remaining two lack logical interpretations and were indeed never observed in our measurements.} of equivalence or difference among \{R1,R2,R3,R4\} and the corresponding interpretation on how the introduced mutation affects the DPI. For example, the scenario \{R1,R3,R4\}\{R2\} indicates that applying the mutation does not alter the web server's behavior (since R1=R3=R4), but effectively prevents the DPI from parsing and recognizing the blocked domain, since the censorship behavior is only observed in Reference Test (R2) but not Mutated Test (R4). In that scenario, \analyzer assigns a ``Bypass'' verdict for the (\textit{Target, Probe}) pair.

\input{tables/analyzer}

In rare cases, the four results may yield no conclusive verdict. Consider the example where R1 (Reference Control) returns a valid HTTP response, but R2, R3, and R4 all appear blackholed (\{R1\}\{R2,R3,R4\}). The apparent similarity between the Reference Test (R2) and the Mutated Control (R3) suggests that the web server itself might be discarding the mutated request, making it indistinguishable from a DPI-induced packet drop. Consequently, we cannot determine if the DPI has been triggered in R4. \analyzer labels such cases as \textit{Inconclusive}. In Appendix \S~\ref{sec:inconclusive} we discuss some approaches we take to reduce the fraction of inconclusive results.

Finally, \analyzer outputs a \textit{fingerprint} for each \textit{Target} by concatenating verdicts across all evaluated \textit{Probes}, with 0 indicating the mutation of the current \textit{Probe} did not affect the DPI and 1 indicating the evaluated mutation disrupted the DPI's ability to interpret traffic. For inconclusive results, \analyzer marks -1 in the fingerprint and omits that probe from pairwise fingerprinting matching.

\subsection{Probe Selection}
\label{sec:probeselection}

\subsubsection{\textbf{Generate Candidate Probes}}
\label{sec:generateprobes}

To generate candidate probes for fingerprinting, we follow a \textit{deterministic, grammar-aware fuzzing} approach guided by known parsing ambiguities identified from prior work (\S~\ref{sec:evasionsurvey}). There are two main reasons for this design choice: (1) Randomly mutating packets as raw bytes (\ie, grammar-agnostic fuzzing) often produces packets that fail basic validity checks and are likely discarded by intermediary routers before they ever reach the DPI. By contrast, grammar-aware fuzzing respects the essential format and semantics of its underlying protocols, generating probes that are more likely to trigger meaningful divergences in DPI behaviors rather than being prematurely discarded. (2) We systematically enumerate precisely which fields to mutate, as well as the range of mutating values (inspired by Table~\ref{tab:evasionsurvey}), such that the \textit{same set of candidate probes} is deterministically applied across all tested DPIs. This allows feature vectors of the resulting \textit{fingerprints} to be directly comparable for clustering or differentiation.

We begin with a standard packet sequence that serves as a baseline template, which consists of 1) a client-initiated SYN packet (with typical TCP options expected from a Linux client like windowscale and SACK), 2) a corresponding ACK for the server's SYN-ACK (with appropriate ACK number), 3) an HTTP GET request or a TLS Clienthello that resembles one produced by cURL, and finally 4) a FIN/ACK and 5) an ACK to gracefully terminate the connection. On top of this baseline, we broadly define three types of mutations:

\textbf{Insertion} is a sequence-level mutation that builds and injects an entirely new packet at a specified point in the baseline sequence. The inserted packet is crafted with systematically varied fields at the IP, TCP, and application layers. For example, we vary TCP flags, SEQ and ACK numbers, checksums, and other header fields. The inserted packet can contain no payload, random bytes, a well-formed application-layer request with the Control Domain, along with others. Each inserted packet can be placed at one of four defined positions: before the initial SYN, before SYN-ACK, before the ACK completing the handshake, or after the handshake but immediately preceding the application-layer request.
    
\textbf{Mutation} refers to packet-level mutations that affect an individual packet (either header or payload) from the baseline sequence. We define mutations for most of the header fields for the IP and TCP layer, except for the fields that are essential for packet routing and delivery, such as IP addresses or TCP ports. For HTTP, being a more expressive text-based protocol, we leverage previously studied HTTP-specific mutations~\cite{raman2022cenfuzzsurvey7, nourin2023measuringsurvey10, harrity2022getsurvey11, goodbyedpisurvey21, yadav2018lightsurvey27}, such as varying the case of HTTP methods (\eg, GET vs. GeT). The full list of IP and TCP fields considered for mutations is included in Appendix~\ref{sec:tcpipmutations}.

\textbf{Fragmentation} is a multi-packet mutation that can be applied as IP fragmentation, TCP segmentation or TLS record-layer fragmentation, which we collectively term as ``fragmentation''. Fragmentation-based mutations control (1) the exact offset where fragmentation occurs (\eg, splitting a domain name across fragments); (2) the order in which fragments are sent (in-order vs. out-order); (3) the size of individual fragments; (4) the total number of fragments created from the original payload; (5) the time delay between sending consecutive fragments (reassembly timeouts). For IP fragmentation in particular, we also consider ``disorder'' fragments, where fragments from different reassembly queues become interleaved (\ie, two sets of fragments sharing the same IPs but carrying different IPIDs).

\input{figures/overlapping}

A major source of ambiguity in fragmentation handling relates to overlapping fragments---that is, two fragments containing different data intended for the same (or partially overlapped) offsets in the reassembled packet. For this, we define the following mutation procedure: first, we build and serialize two byte sequences from the same application-layer request, with sequence A using the original domain (\ie, the Control or Test Domain) and sequence B using the same domain but reversed in 16-bit chunks. (Because TCP checksums are calculated using one's-complement on 16-bit chunks, the second sequence, with only domain reversed and everything else being the same, remains a semantically valid packet with correct checksums.) Note that this reversed domain is likely malformed and therefore unlikely to match any entry in the DPI's blocklist. Next, we build fragment $X$ that span the byte range [$X_L$, $X_R$] from sequence A, and fragment $Y$ that span [$Y_L$, $Y_R$] from sequence B. We ensure that the overlapping portion covers the entire domain name region so that the fragments ``equivocate'' over exactly which domain is used. As shown in Figure~\ref{fig:overlapping}, we define mutations corresponding to each of the nine possible alignments of fragment boundaries ($X_L$, $X_R$, $Y_L$, $Y_R$). Finally, we send the fragments and observe whether the DPI is triggered, which allows us to infer which domain is present in the reassembled packet and, in turn, how the DPI handles overlapping in fragmentation reassembly.
    
During probe generation, we may allow our fuzzer to apply up to $N$ mutations per probe. We note that even with $N=1$, this process yields 2,621 HTTP-based probes and 2,590 HTTPS-based probes, which, as we show in the next subsection, already introduce substantial variance in DPI behaviors. One-mutation-per-probe also simplifies root-cause analysis (\S~\ref{sec:rootcause}) by isolating the effect of an individual mutation. Therefore, in this work, we restrict our probe generation to single-mutation probes ($N=1$). We note that single-mutation already covers the vast majority of known evasion attacks cataloged in our evasion survey.

\input{algos/probeselection}

\subsubsection{\textbf{Filter and Select Probes}}
\label{sec:filterprobes}

From the pool of candidate probes, we aim to filter and select those that are most \textit{discriminative}---\ie, those that elicit the greatest diversity of behaviors across different DPI implementations. Doing so requires access to a set of \textit{known} DPIs. Acquiring such a set, however, proved one of the most challenging tasks for this work, as most DPI hardwares are costly and often require separate licenses to enable key filtering features. For this, we ultimately assembled our set of known DPIs from three complementary sources: \textbf{Open source DPIs (4)}: We configured four popular open-source DPIs (zeek, nDPI, Suricata, and Snort). \textbf{Commercial free trials (3)}: We also deployed three leading commercial firewalls offered as free trials through the AWS Marketplace (Cisco Secure Firewall, Fortinet FortiGate, and Sophos UTM 9) in a Virtual Private Cloud. \textbf{DPIs with identifying blockpages (11)}: Finally, we leveraged public measurement data from the Censored Planet Observatory~\cite{sundara2020censored}. Over one week in February 2025, Censored Planet recorded measurements to 251 remote endpoint addresses showing injected blockpages, of which 95 contain vendor-identifying information that is attributable to 11 distinct DPI vendors. We then tested the pool of candidate probes on these known DPIs.

\textit{Pre-filters.} We begin by discarding any probes that consistently yield uniform responses across all tested DPIs. Such probes offer no discriminative power, often because their mutations are either too trivial that have no observable effect (\eg, inserting an empty ACK after handshake) or too disruptive that all DPIs discard them (\eg, mutating the \textit{proto} field in the IP header). Next, we remove probes whose outcomes frequently ($\geq 10\%$) lead to inconclusive interpretations, as defined earlier in Table~\ref{tab:analyzer}. These inconclusive probes provide limited discriminative value. Through this pre-filter step, we reduce the pool of candidate probes by around 70\% (2,621 to 702 for HTTP, and 2590 to 708 for HTTPS).

\textit{Ranking probes by Entropy.} From here, our selection of probes conceptually followed a ``Maximum Entropy, Minimum Redundancy'' approach, as outlined in Algorithm~\ref{alg:probeselection}. We begin by treating the outcome of each probe as a binary feature, and measure its capacity to separate the known DPIs by calculating the Shannon entropy of each probe's distribution of \textit{Bypass}/\textit{NoEffect} across all distinct DPI groups. Intuitively, a probe that splits distinct DPIs roughly evenly (\eg, half the DPI groups bypassed, half not) has higher entropy and is thus more discriminative. We rank the remaining candidate probes in descending order of their entropy.

\textit{Greedy probe selection with correlation check.} Next, we apply a greedy selection algorithm (Algorithm~\ref{alg:probeselection}) in which we iteratively select the highest-entropy probe and compute its phi coefficient with each probe \textit{already} in the selected set. We only add the new probe if its minimum pairwise correlation with the existing set is below a threshold (we used $\phi=0.85$, determined empirically). This ensures that each newly added probe contributes new information rather than duplicating the effect of a probe already in the set (\eg, if two mutations are closely related). For example, two probes might both divide the DPIs into two equal halves (thus both having the highest entropy), but if they partition the groups identically, adding both would be redundant. \textbf{Table~\ref{tab:selectedprobes} in Appendix lists the top 40 probes selected from this process for HTTP and HTTPS.}

\input{figures/distance_N}

\textit{Number of probes used.} Finally, a decision needs to be made regarding how many of the selected probes to include in real-world measurements. Figure~\ref{fig:distance} shows that with as few as ten probes, our current DPI set is already fully distinguishable---though the set is limited in its size. As additional probes are added, the minimum and average Hamming distance among DPIs continue rising nearly linearly. With 20 probes, the average pairwise distance is approximately ten bits, meaning about half of the fingerprint bits differ between any two DPIs. Eventually the curve plateaus after 30-40 probes, where the marginal benefit of adding more probes begins to diminish. In practice, the number of probes must balance improved discrimination against the increased measurement overhead: each probe run (including Control \& Test, and a conservative wait in-between) requires around 140 seconds in our setup, so more probes per target constrains the scale of targets can feasibly be tested.

\input{figures/cpcluster}

As an example, Figure~\ref{fig:cpcluster} presents a two-dimensional Multidimensional Scaling (MDS) plot based on 20‑bit fingerprints produced by our top‑20 probes. Each point represents one endpoint (from the Censored Planet dataset) for which an in-path DPI injects a vendor‑identifying blockpage, with color-coding by vendor. The axes represent a two-dimensional projection whose Euclidean distances approximate the Hamming distances among fingerprints. We observe that endpoints associated with the same blockpage vendor generally cluster together, yet each cluster also exhibits some internal spread. This variation likely reflects the reality of remote measurement where additional in‑path network devices (\eg, other middleboxes) may modify traffic before it reaches the DPI. We discuss this limitation further in \S~\ref{sec:limitation}. Despite these noise factors, most endpoints associated with the same vendor’s blockpage still end up meaningfully closer to each other than to endpoints associated with different vendors.

\subsubsection{\textbf{Root Cause Analysis}}
\label{sec:rootcause}

An advantage of using single-mutation probes is that each probe can often be traced to a specific ambiguity in traffic interpretation. Coupling this with open-source DPIs affords us an opportunity to understand \textit{why} certain probes discriminate among DPIs. For example, one of our top-ranked probes (listed in Table~\ref{tab:selectedprobes}) mutates the sequence number ($SEQ$) of the packet with the triggering request. Specifically, it sets the $SEQ$ to a negative value relative to the client's initial sequence number, which places the $SEQ$ outside the receiver's window, but also prepends the payload with padding bytes to align the portion of payload containing the request exactly at the next expected sequence number.

\input{probes/63}

This probe triggered a censorship response from Snort (v3.6.0), but failed to trigger a response from Zeek (v7.0.4). Inspecting their source code (Figure~\ref{fig:zeek63},~\ref{fig:snort63} in Appendix), we found that while both DPIs discard packets with invalid $SEQ$, their definitions of ``invalid'' diverge. Zeek simply compares the current $SEQ$ to the client's initial sequence number; if the current $SEQ$ is lower, Zeek labels the packet as ``seq\_underflow'' and considers its payload invalid for reassembly. In contrast, Snort implements a more nuanced validation: it checks whether the current $SEQ$ falls below the upper boundary of the receiver's window ($seq \leq rcv\_next + window\_size$), and also whether the last byte of the payload is above the lower boundary of the receiver's window ($seq + len(packet) \geq rcv\_next$). Note that there is no lower bound for $SEQ$ to be valid. As such, the part of the payload that is in-window is then processed by Snort for flow reassembly, and eventually triggers its censorship policy. In this case, the subtle mismatch in how Zeek and Snort handle partially out-of-window segments enables a remote prober to distinguish the two implementations.

\input{probes/55}

Another example involves mutating the TCP timestamp ($TSval$) of the request packet. Specifically, the probe selects a $TSval$ lower than that of the preceding packet. For Zeek and nDPI, this mutation has no impact---both DPIs continue to process the packet, ultimately triggering a censorship response. Snort, however, discards the packet, so no censorship response is observed.

Snort’s behavior stems from its implementation of ``Protection Against Wrapped Sequences'' (PAWS), a mechanism defined in RFC7323~\cite{rfc7323} that discards segments if their $TSval$ is smaller than timestamps recently observed on the same connection. The fact that Zeek and nDPI do not implement this check allows them to be distinguished from Snort. Interesting, Snort's implementation (Figure~\ref{fig:snort55} in Appendix) also deviates from the RFC's specification: the RFC states that if a connection has been idle long enough, the result from the PAWS check should be ``invalidated'' (\ie, the packet should be accepted). Snort developers, however, interpreted ``invalidate'' to mean invalidating (discarding) the packet itself rather than the PAWS check, resulting in the \textit{exact opposite} of RFC-intended behavior. We have reported this issue to Snort developers. This example reinforces how, even when implementations strive for RFC compliance, differences in implementers' interpretations can introduce ambiguities that ultimately enable fingerprinting.

%% file: figures/cendpi.tex
\begin{figure*}[t]
\centering
 \includegraphics[width=2\columnwidth,keepaspectratio]{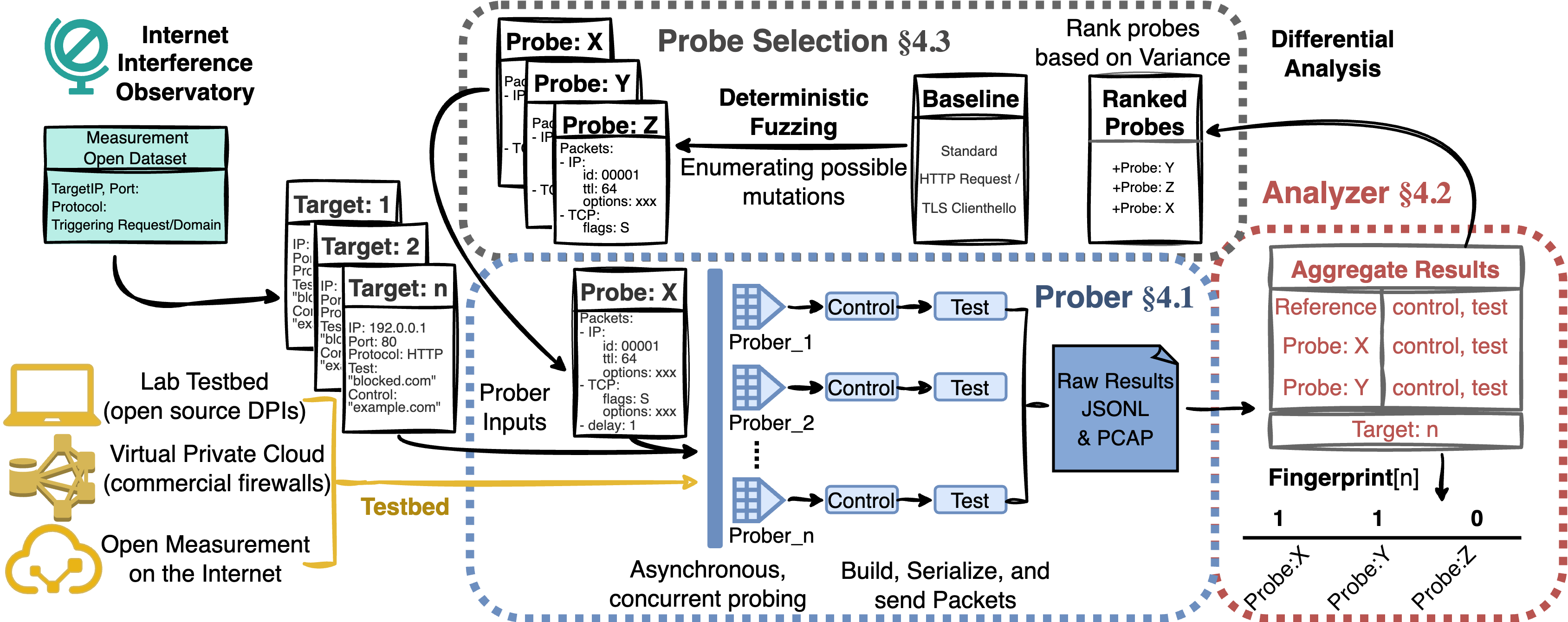}
%\vspace{-5pt}
\caption{\nameofthething's Architecture. The framework sources from open measurement datasets for target (web) servers behind DPIs, sends probes that are enumerated and ranked based on differential analysis, and analyzes results to produce DPI fingerprints.}
\label{fig:cendpi}
%\vspace{-10pt}
\end{figure*}

%% file: tables/analyzer.tex
\begin{table}[t]
\centering
\resizebox{1.03\columnwidth}{!}{
\begin{tabular}{p{0.2\columnwidth}p{0.92\columnwidth}p{0.15\columnwidth}}
\toprule
\textbf{Results} & \textbf{Interpretation} & \textbf{Verdict} \\ \hline
\midrule

\{R1,R3,R4\}\{R2\} & Mutation has no effect on endhost but bypasses DPI & \textbf{Bypass} \\
\{R1\}\{R2,R3,R4\} & Reference blocking behavior indistinguishable from how endhost handles mutation, unclear if DPI is triggered in R4 & Inconclusive \\

\{R1,R3\}\{R2,R4\} & Mutation has no impact on either the DPI or the endhost & \textbf{NoEffect} \\

\{R1,R3\}\{R2\}\{R4\} & Mutation does not affect endhost; however, differing blocking behaviors imply possibly two distinct DPIs in the path & Inconclusive \\

\{R2,R3\}\{R1\}\{R4\} & Possibly two DPIs in path: first drops all packets when triggered (R2), superseding second DPI's explicit blockpage. Mutation bypasses the first DPI but not the second.  & Inconclusive \\

\{R2,R4\}\{R1\}\{R3\} & Mutation affects the endhost but the DPI remains triggered. & \textbf{NoEffect} \\

\{R3,R4\}\{R1\}\{R2\} & Could happen when Mutated Control (R3) and Mutated Test (R4) both yield no response, different from reference behaviors. & Inconclusive \\

\{R1\}\{R2\}\{R3\}\{R4\} & Mutation affects both the endhost and potentially multiple DPI implementations in the path that respond differently. & Inconclusive \\

\bottomrule
\end{tabular}
}
\caption{\small Interpretation of probe outcomes. Each row corresponds to a unique grouping of responses. $\{a,b\}\{x,y\}$ indicates $a=b\ \&\ x=y\ \&\ a \neq x$. Some invalid groupings (\eg, \{R1,R2\}) are excluded.}
\label{tab:analyzer}
%\vspace{-9mm}
\end{table}

%% file: figures/overlapping.tex
\begin{figure}[t]
\centering
 \includegraphics[width=\columnwidth,keepaspectratio]{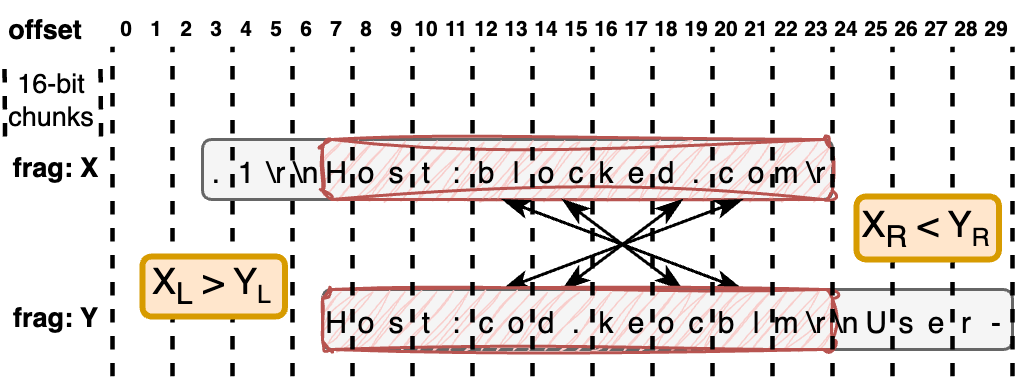}
%\vspace{-15pt}
\caption{Overlapping Fragment Reassembly. Nine unique alignments occur based on whether $Y_{L}$/$Y_{R}$ are smaller than, equal to, or greater than $X_{L}$/$X_{R}$. Figure inspired by~\cite{khattak2013towardssurvey6}.}
\label{fig:overlapping}
%\vspace{-10pt}
\end{figure}

%% file: algos/probeselection.tex
\begin{algorithm}[t]
\caption{Probe Selection with Entropy and Greedy Correlation}
\footnotesize
\label{alg:probeselection}
\begin{algorithmic}[1]
\Require
  \Statex $\mathcal{P}$: Set of all candidate probes; $\mathcal{D}$: Set of known DPIs (groups)
  \Statex $f(p,d)\in\{\textit{Bypass},\textit{NoEffect},\textit{Inc}\}$:
           outcome of probe $p$ on DPI $d$;
  \Statex $\theta$: correlation threshold
\Ensure
  \Statex $\mathcal{S}$: Final subset of selected probes

\State \textbf{for each} $p\in\mathcal{P}$, compute $\mathrm{score}[p] \gets \textsc{getEntropy}(p,\mathcal{D},f)$
\State $\mathcal{P}_\mathrm{sorted}\gets\textsc{sort}(\mathcal{P}\text{, by }\mathrm{score}[p]\text{ descending})$; $\mathcal{S}\gets\emptyset$
\For{$p$ in $\mathcal{P}_\mathrm{sorted}$}
  \State $\textit{keep}\gets \text{true}$
  \For{$q \in \mathcal{S}$}
    \If{$|\textsc{getPhiCoefficient}(p,q,\mathcal{D},f)|  > \theta$} 
      \State $\textit{keep}\gets\text{false};$ \textbf{break}
    \EndIf\EndFor
  \If{\textit{keep}} $\mathcal{S}\gets \mathcal{S}\cup\{p\}$ \EndIf
\EndFor
\State \Return $\mathcal{S}$

\end{algorithmic}
\end{algorithm}

%% file: figures/distance_N.tex
\begin{figure}[t]
\centering
 \includegraphics[width=\columnwidth,keepaspectratio]{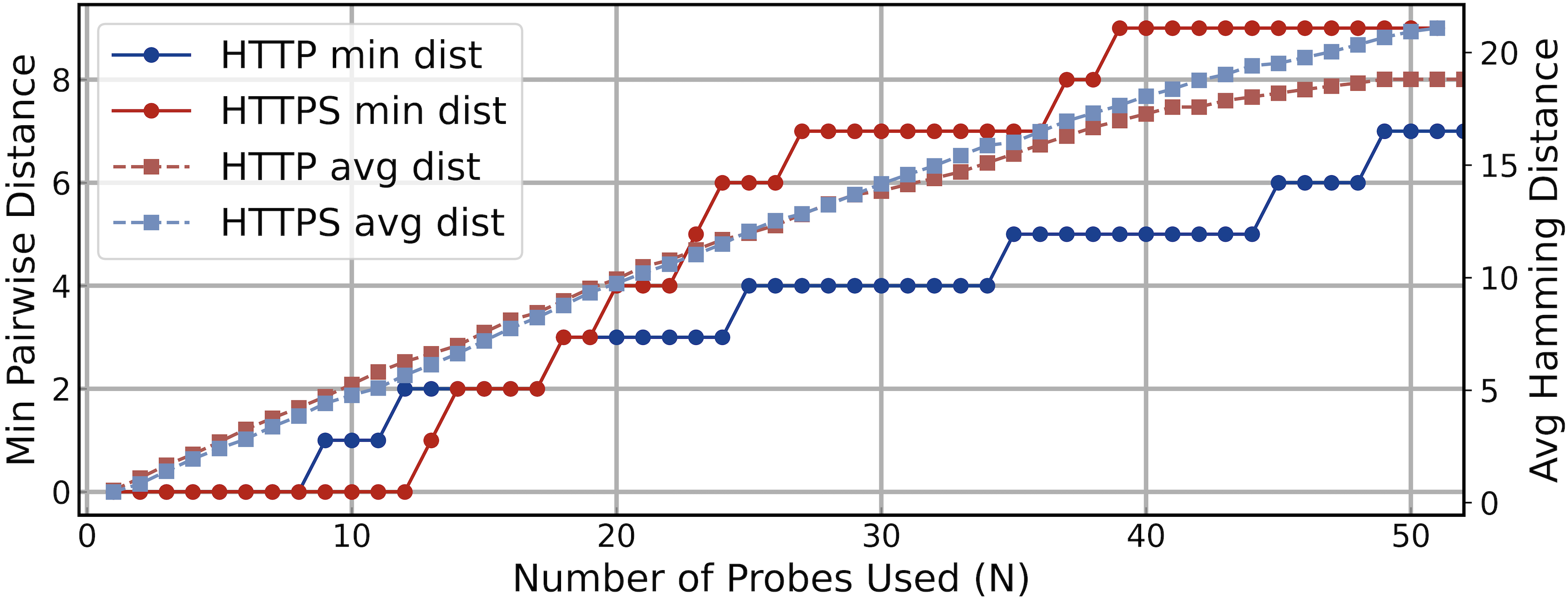}
%\vspace{-15pt}
\caption{Minimum and average pairwise Hamming distances among known DPIs from different vendors, plotted against the number of top N probes used.}
\label{fig:distance}
%\vspace{-15pt}
\end{figure}

%% file: figures/cpcluster.tex
\begin{figure}[t]
\centering
 \includegraphics[width=\columnwidth,keepaspectratio]{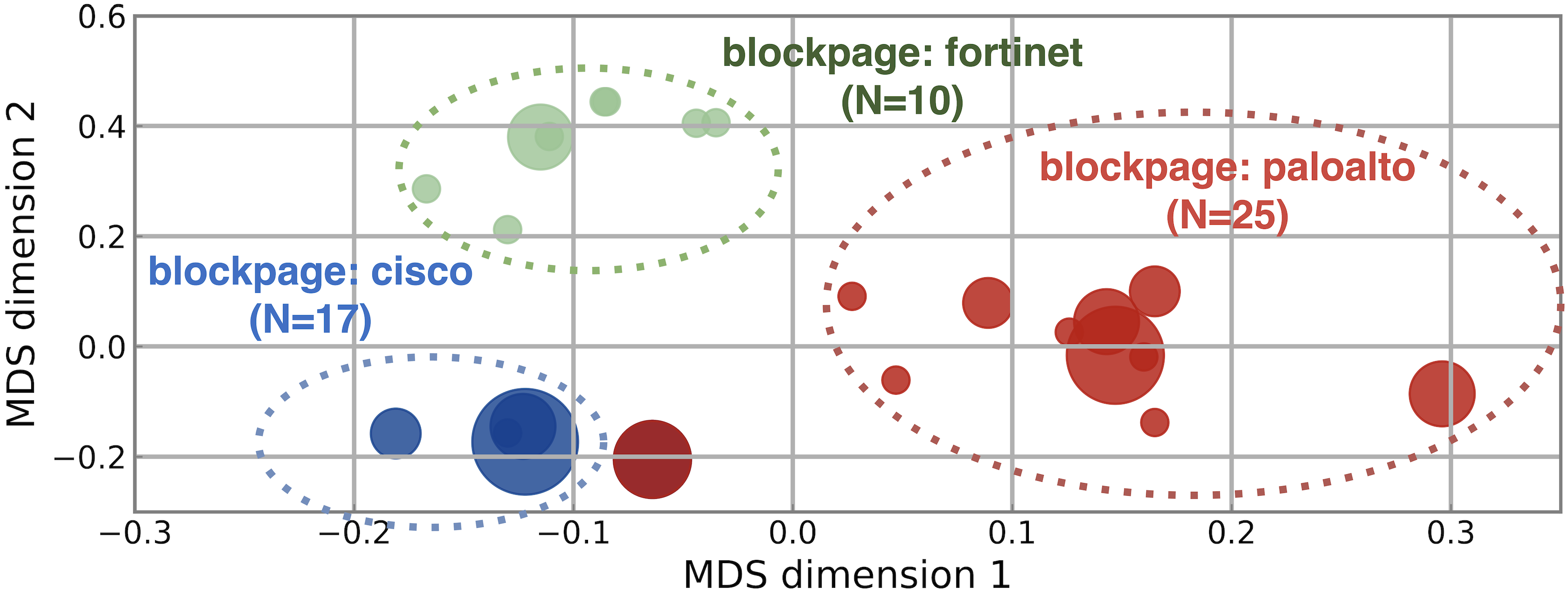}
%\vspace{-15pt}
\caption{MDS of 20-probe fingerprints using Hamming distance. Each is an endpoint that, when probed, triggered vendor-identifying blockpages by DPI; colored by vendor.}
\label{fig:cpcluster}
%\vspace{-15pt}
\end{figure}

%% file: probes/63.tex
\begin{table}[h]
\small
\centering
\begin{tabular}{p{0.97\columnwidth}}
\toprule
\textbf{Mutate\{layer:TCP;field:seq;option:negativeSeqWithPadding\}}
\\
\toprule

$packet_0$: [SYN \space (seq=$ISN_{client}$, ack=0)]\\
\textbackslash \textbackslash \space waiting for incoming ([SYNACK \space (seq=$ISN_{server}$, ack=$ISN_{client}$+1)])\\
$packet_1$: [ACK \space (seq=$ISN_{client}$+1, ack=$ISN_{server}$+1)]\\
$packet_2$: [PSH/ACK \space (seq=$ISN_{client}$-100, ack=$ISN_{server}$+1, \\
\hspace{25mm} payload=[0]*101 + request[TestDomain])]\\
$packet_{3-4}$: [FIN/ACK...], [ACK ...] \\

\bottomrule
\end{tabular}
\captionsetup{labelformat=empty}
\caption{}
\label{tab:probe63}
\vspace{-10mm}
\end{table}

%% file: probes/55.tex
\begin{table}[h]
\small
\centering
\begin{tabular}{p{0.97\columnwidth}}
\toprule
\textbf{Mutate\{layer:TCP;field:option;option:timestamp\}}
\\
\toprule

$packet_0$: [SYN \space (timestamp: 1000)]\\
\textbackslash \textbackslash \space waiting for incoming ([SYNACK \space (timestamp: 2000)])\\
$packet_1$: [ACK \space (timestamp: 1001)]\\
$packet_2$: [PSH/ACK \space (timestamp: 999, payload=request[TestDomain])]\\
$packet_{3-4}$: [FIN/ACK...], [ACK ...] \\

\bottomrule
\end{tabular}
\captionsetup{labelformat=empty}
\caption{}
\label{tab:probe55}
\vspace{-10mm}
\end{table}

%% file: 05measurement.tex
\section{Measurement}
\label{sec:measurement}

We applied \nameofthething for large-scale measurements to fingerprint in-path/on-path DPI devices filtering HTTP and HTTPS traffic, based on our methodology developed in the previous sections.

\subsection{Measurement Setup}

All measurements were performed from a dedicated measurement machine hosted in North America, administered by the research department of a regional, education-focused ISP. Because some of our probes mutate IP-layer fields, we took steps to ensure that egress traffic was minimally normalized by the local network environment. For example, we verified that no local middlebox or router was reassembling IP fragments before they leave our network\footnote{We note that even if there had been filtering that neutralized specific mutations (\eg, reassembling all outbound IP fragments), it would naturally emerge during probe selection as a uniform response across all tested DPIs. Such probes would then be discarded due to low discriminative power.}. We began our measurements in early February 2025.

\textbf{Input list of \textit{Probes}} We drew on the procedure described in \S~\ref{sec:probeselection} to select the top 40 probes for HTTP and HTTPS. Of the 40, 21 probes were common to both protocols. Table~\ref{tab:selectedprobes} in Appendix lists these probes along with descriptions of their specific mutations.

\textbf{Input list of \textit{Targets}} Our fingerprinting measurements incur a non-trivial time overhead---around 140 seconds for each of the 40 selected probes. This makes it infeasible to blindly test the entire IPv4 space with \nameofthething directly. Instead, we supplied \nameofthething with a curated list of \textit{Targets} already known to have DPIs interfering with HTTP/HTTPS traffic along the network path. We gathered these \textit{Targets} from open measurement data provided by the Censored Planet Observatory from February 2025~\cite{sundara2020censored}. Specifically, we took every IP/port combination where blocking (\eg, RST injection) was detected for at least one domain tested by CP. If multiple domains triggered censorship for the same target, we randomly selected one to be used as the Test Domain for \nameofthething measurement. In total, this produced 11,467 unique \textit{Targets} for HTTP and 22,092 for HTTPS, spanning 482 network prefixes (based on the Routeviews~\cite{Routeviews} dataset from CAIDA), 179 ASes, and 73 countries. We probed each target three times using each of the 40 selected probes, launching over 3 million measurements in total. We describe the ethical considerations in our measurement design in Appendix~\ref{sec:ethics}.

\subsection{Measurement Results}

\subsubsection{Fingerprint Clustering at Network, AS, and Country Levels}

\input{figures/pariwise_distance}

We first examined the pairwise similarity between targets’ 40-bit DPI fingerprints, focusing on how this varies at different network scopes. Figure~\ref{fig:pairwisedistance} shows the distribution of pairwise Hamming distances for all target pairs and compares it with the same distribution for target pairs within the same netblock, Autonomous System (AS), or country. The overall distribution (leftmost) is quite spread, indicating the overall diversity of DPI behaviors elicited by our probes. However, looking at narrower network scopes, targets from the same netblock or AS generally have more similar fingerprints. For example, only $< 1\%$ of all target pairs have a Hamming distance of 0, but this fraction jumps to 44.97\% and 36.09\% (HTTP/HTTPS) among pairs in the same AS, and to 52.60\% and 52.33\% among pairs in the same netblock. That means a substantial fraction of pairs sharing the same netblocks also share the exact identical 40-bit fingerprint, suggesting that censorship policies at a given network or AS level are often enforced by the same DPI devices or at least highly similar implementations.

\input{figures/country_cluster}

Next, we applied HDBSCAN~\cite{hdbscan}, a hierarchical density-based clustering method, to group targets by their 40-bit DPI fingerprints. Unlike many clustering algorithms, HDBSCAN does not require specifying the number of clusters a priori; instead, it identifies ``core'' regions of density automatically, an advantage when we do not know the variety of DPI implementations in the wild. In total, we find 203 clusters that have at least 20 targets. Figure~\ref{fig:countrycluster} shows the top discovered clusters: the majority of clusters indicate localized deployment of a particular DPI product or configuration that is prevalent in just one region, while some span multiple countries---possibly reflecting more globally distributed commercial solutions adopted by different networks. For HTTP measurements returning blockpages, we cross-referenced the responses with Censored Planet’s database of known blockpages to corroborate our results. For example, 100\% of targets in clusters \#8 and \#23 are located in Iran and emit a blockpage identified with Iran’s national firewall~\cite{lange2025ra, bock2020detectingsurvey15}, while 97\% of targets in cluster \#63 produce blockpages consistent with FortiGate firewalls. Notably, targets from cluster \#63 are spread across 13 different countries, yet they share very similar 40-bit fingerprints due to (presumably) the similarity in their underlying DPI implementations.

A more unexpected finding emerged from our measurements in China: instead of forming a single monolithic cluster, Chinese targets split into multiple clusters (\eg, \#178, \#179, \#200, \#170, and \#183). Although prior research often characterizes the Great Firewall of China (GFW) as highly uniform, our observation indicates some variability among different Chinese ASes. At the netblock level, this separation is even more clear---most CN netblocks appear in only one cluster each. This result suggests that while the \textit{effect} of censorship (\ie, which sites are blocked) may remain consistent across the country, the \textit{implementation} of censorship may be less homogeneous than previously assumed. These discrepancies could reflect different versions of the DPI infrastructure colloquially known as the ``GFW'', or the existence of additional middleboxes deployed at the regional or provincial levels. Regardless, these observations clearly challenges the conventional view of the GFW as a singular, homogeneous firewall.

%\subsubsection{The effect of fingerprint length on clustering}

\input{figures/http_https}
\subsubsection{HTTP \& HTTPS fingerprints} Our fingerprinting probes include 21 TCP/IP-layer probes that are shared across and agnostic to the application-layer protocol. Using these 21 shared probes, we compare the partial fingerprints for targets tested under both protocols. We filtered our results to retain only those target IPs for which we obtained valid fingerprints in both HTTP and HTTPS measurements, and we further restricted our analysis to include only targets measured using the \textit{same Test Domain} for both protocols to reduce the possibility of triggering different DPIs with different domains. In total, we identified 4,147 unique targets (``common targets'') for which we have paired HTTP and HTTPS fingerprints.

Figure~\ref{fig:httphttps} shows the distribution of Hamming distances between HTTP and HTTPS fingerprints across the 4,147 ``common targets'', using the 21 shared bits. For comparison, we also plot baseline distribution of distances between randomly paired fingerprints from different IPs. The results reveals that fingerprints at the TCP/IP layer remain highly consistent for the same target across the two protocols: $\geq 50\%$ of ``common targets'' exhibit identical, protocol-agnostic fingerprints, while almost $80\%$ differ by at most one bit. This level of similarity is significantly higher than the baseline random pairing, indicating that the DPI implementations filtering HTTP and HTTPS traffic along the same network path are typically identical (\ie, same device) or, at minimum, share substantial commonality in their TCP/IP-layer processing logic.

\subsubsection{Fingerprinting in noisy environments}
\label{sec:cnnoise}
In examining per-country fingerprint clustering, we noticed that targets from China (CN), Turkey (TR), and Cuba (CU) exhibited substantially higher pairwise distances within their clusters compared to other countries. A closer look revealed that censorship in these regions often fails to trigger consistently. Even using the same Test Domain and an otherwise identical packet sequence, repeated measurements found different DPI behaviors (\textit{Blocking} vs. \textit{NoBlocking}) 12.20\% of the time for targets in CN and 16.49\% for TR---compared to a global average of 1.68\%. As a result, noise from inconsistent blocking causes ``bit-flipping'' in the measured fingerprints.

\input{figures/repeatprobing}

One way to mitigate such inconsistencies is through repeated measurements. Specifically, we can repeat each probe multiple times, and then aggregate the outcomes by prioritizing any observed ``blocking'' over ``no-blocking'', under the assumption that most DPIs fail-open rather than fail-close (refer to \S~\ref{sec:analyzer}). Figure~\ref{fig:repeatprobing} visualizes the effect of repeated measurements using MDS plots for CN fingerprints, comparing single-shot measurements versus aggregations from three and ten repeated measurements. Clearly, repeated probing significantly reduces measurement noise from sporadic blocking, resulting in tighter and more coherent fingerprint clusters. These results suggest a practical trade-off: measurements must balance the increased overhead from repeated probing against the noise level of the DPI environment being fingerprinted. Encouragingly, our data indicate that the reliability of DPI blocking in most other countries remains high enough to produce stable fingerprints without requiring extensive repetitions.

\input{figures/versions}
\subsubsection{The churn rate of fingerprints over time}
To better understand how DPI fingerprints change over time, we performed repeated measurements on a selected subset of targets over a span of two weeks. For this measurement, we selected a subset of 1,044 HTTPS targets (excluding CN) that presented Extended Validation (EV) TLS certificates---these targets typically belong to large organizations that are more stable over time for repeated probing, and their administrators likely have more resources to examine inbound traffic if needed (although we note our repeated measurements sent less than 40 Kilobytes of traffic per day per target). Each day for 14 days, we applied the same set of 40 probes and compared the resulting fingerprint to that of day one. Across the two-week period, most targets maintained an \textit{identical} 40-bit fingerprint, with only 6.5\% targets' changed by more than a single bit, and another 8.1\% no longer listening on the probed port.

To examine how DPI fingerprints change over a longer timescale, we examined the fingerprint evolution of two popular open-source DPIs---Zeek and Suricata---across their past docker releases. We tested 43 historical versions of Zeek, from version 4.0.0 through the latest version 7.1.0, covering roughly four years, as well as 37 versions of Suricata, from version 5.0.0 through the latest version 7.0.8, spanning over five years. Figure~\ref{fig:versions} plots the cumulative bit-differences in each DPI's fingerprint relative to their earliest tested version. Notably, Zeek's fingerprint remained remarkably consistent across three major releases, indicating that while new features may have been added, its underlying packet-parsing and flow-tracking logic remained largely unchanged. Suricata's fingerprints underwent two noticeable changes over the five-year span, for both HTTP and HTTPS. These results suggest that DPI fingerprints---at least for these open-source implementations---might not change frequently enough as to demand constant re-measurement.

%% file: figures/pariwise_distance.tex
\begin{figure}[t]
\centering
 \includegraphics[width=\columnwidth,keepaspectratio]{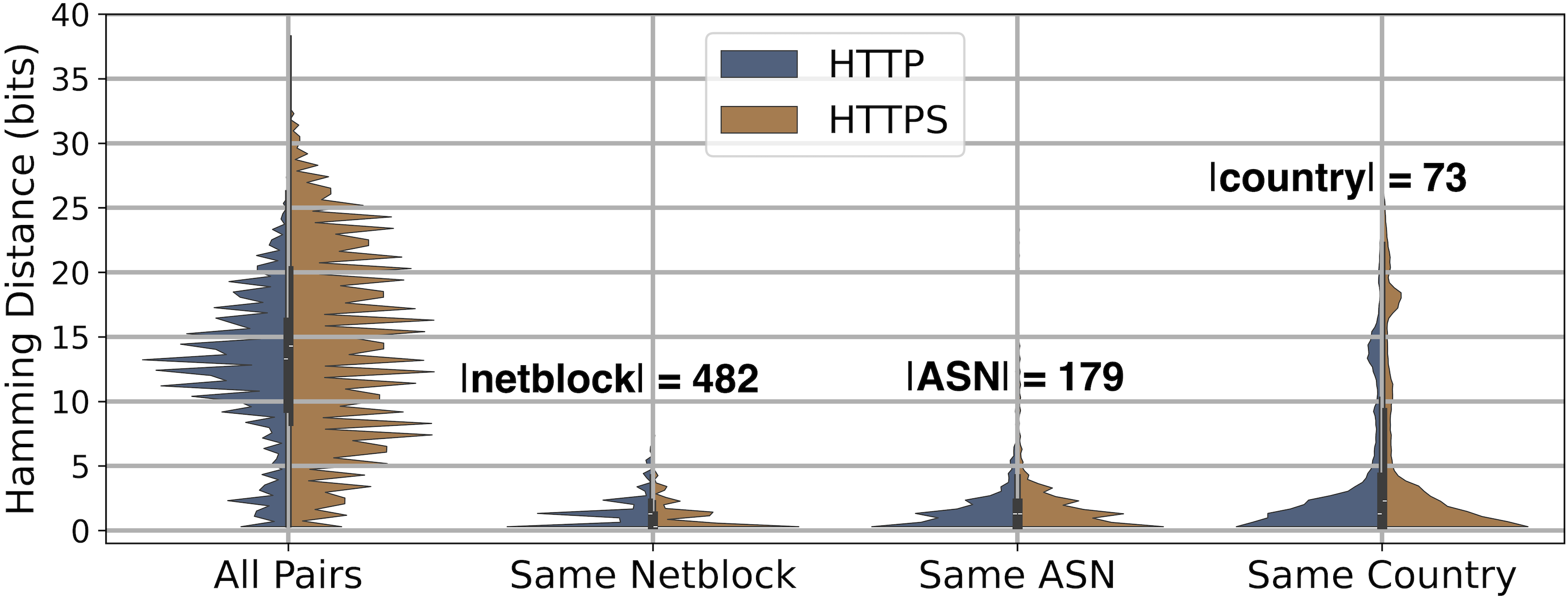}
%\vspace{-15pt}
\caption{Normalized distribution of pairwise fingerprint distances for all pairs of targets, pairs sharing the same netblock prefix, pairs from the same AS, and pairs in the same country.}
\label{fig:pairwisedistance}
%\vspace{-10pt}
\end{figure}

%% file: figures/country_cluster.tex
\begin{figure}[t]
\centering
 \includegraphics[width=\columnwidth,keepaspectratio]{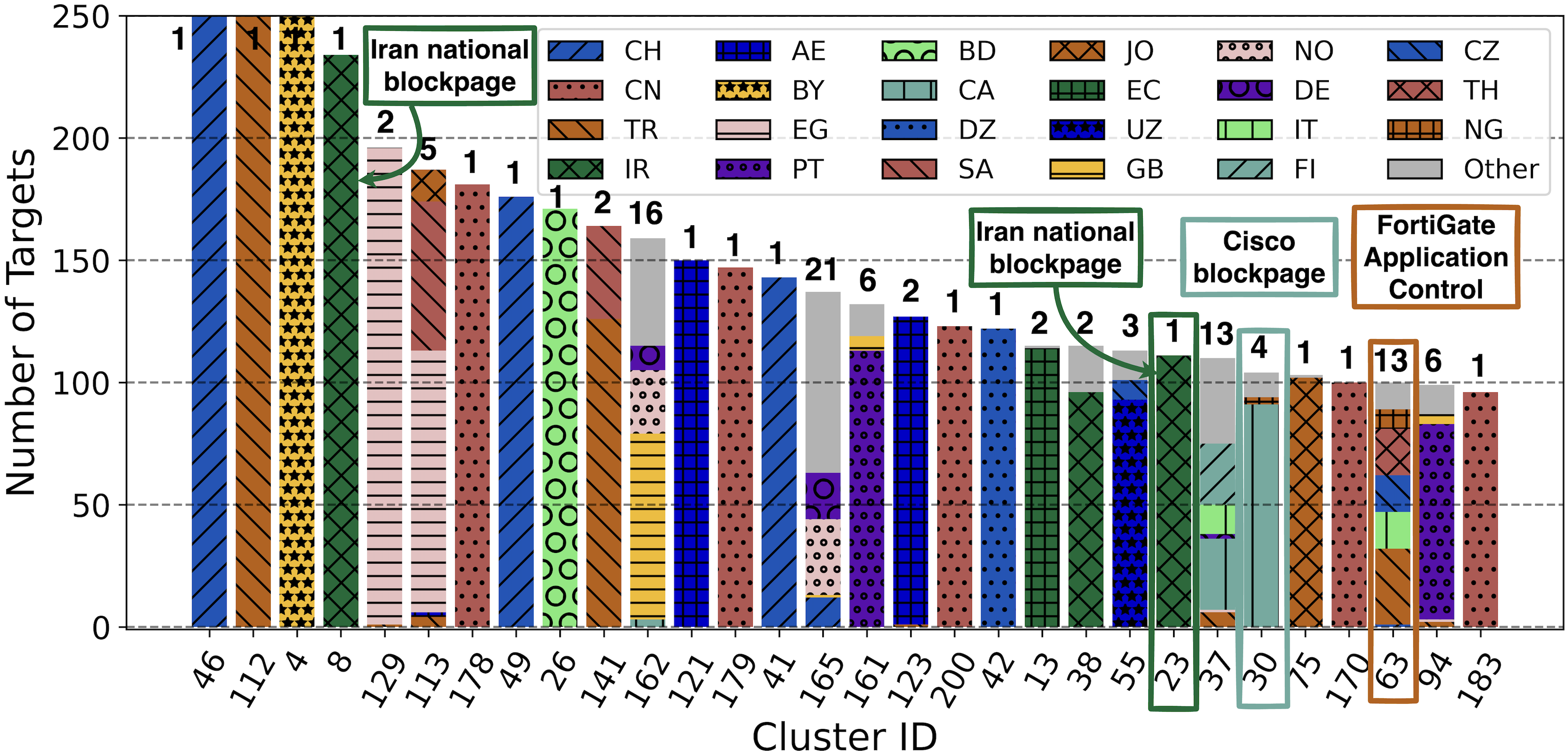}
%\vspace{-15pt}
\caption{Distribution of the top clusters on the 40-bit fingerprints. Bar segments colored by country, with the number on top indicate the number of unique countries in that cluster.}
\label{fig:countrycluster}
%\vspace{-15pt}
\end{figure}

%% file: figures/http_https.tex
\begin{figure}[t]
\centering
 \includegraphics[width=0.85\columnwidth,keepaspectratio]{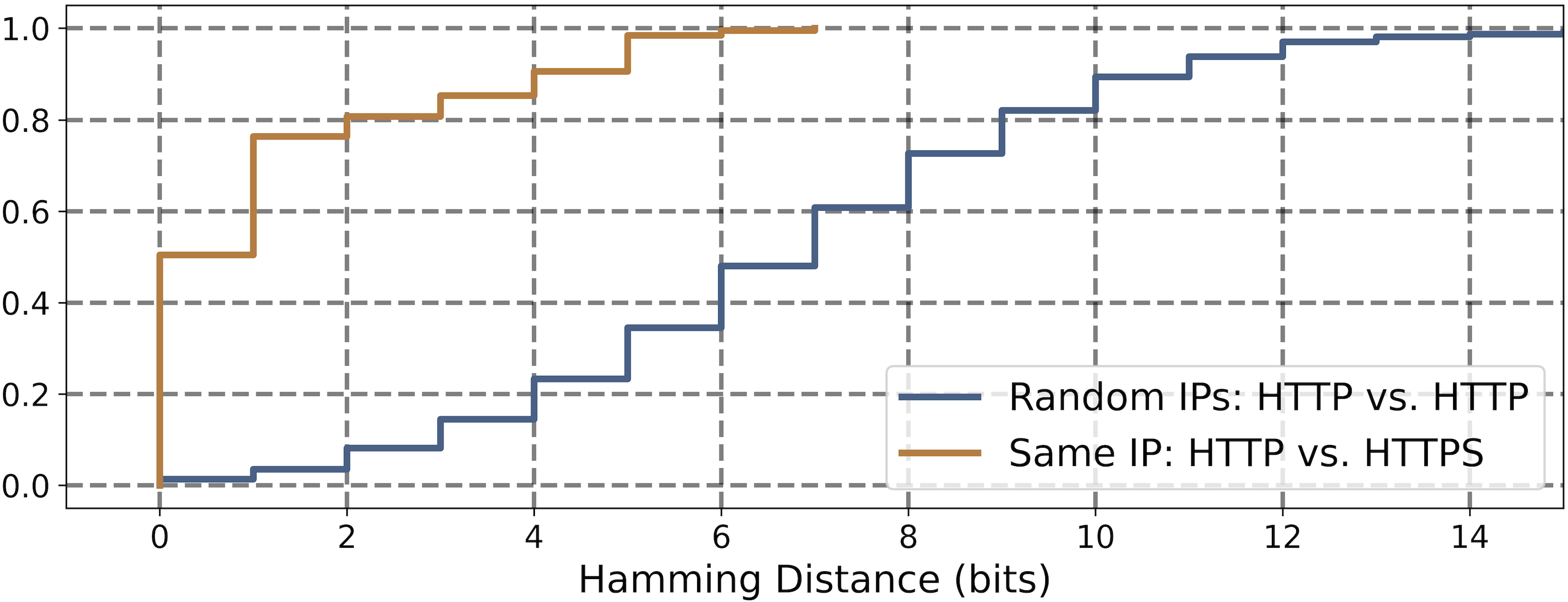}
%\vspace{-10pt}
\caption{Distribution of distances (w/ 21 protocol-agnostic bits) between 1) HTTP and HTTPS fingerprints from same IP vs. 2) randomly paired fingerprints from different IPs.}
\label{fig:httphttps}
%\vspace{-15pt}
\end{figure}

%% file: figures/repeatprobing.tex
\begin{figure}[t]
\centering
 \includegraphics[width=\columnwidth,keepaspectratio]{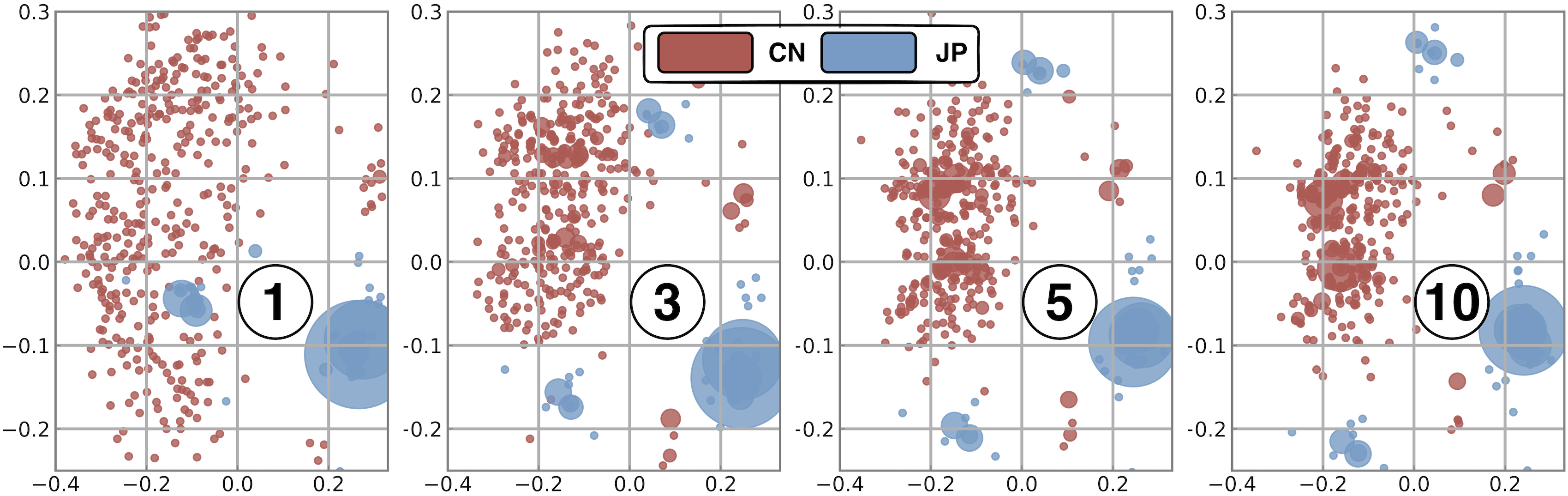}
%\vspace{-15pt}
\caption{MDS plots showing DPI fingerprints for CN targets, obtained from single-shot measurements vs. aggregations from 3/5/10 repeated measurements.}
\label{fig:repeatprobing}
%\vspace{-5pt}
\end{figure}

%% file: figures/versions.tex
\begin{figure}[t]
\centering
 \includegraphics[width=\columnwidth,keepaspectratio]{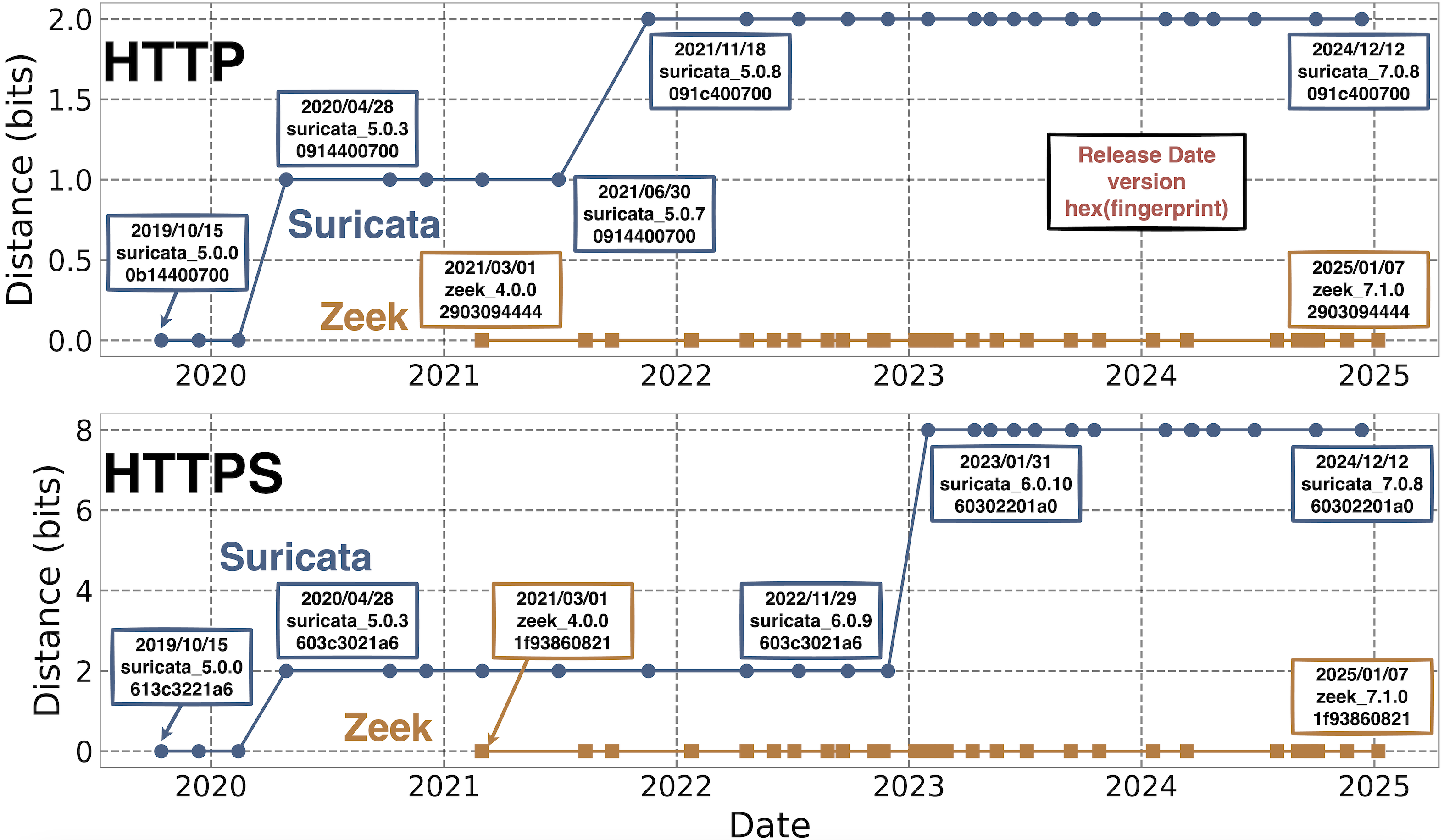}
%\vspace{-15pt}
\caption{Longitudinal changes in their 40-bit fingerprints (hex-encoded) for Suricata and Zeek across version releases.}
\label{fig:versions}
%\vspace{-15pt}
\end{figure}

%% file: 06discussion.tex
\section{Discussion}
\label{sec:discussion}

\subsection{Long-Term Viability of DPI Fingerprinting} 
One question we ask ourselves is whether the proposed DPI fingerprinting methodology will remain effective in the long term. Is the variance in DPI behaviors that enables fingerprinting merely a transient phenomenon of buggy implementations, or will this variance persist? Reflecting on lessons learned from the history of TLS fingerprinting~\cite{Frolov2019a}, we see two potential scenarios under which DPIs could cease being fingerprintable under our methodology, and we discuss why neither is likely to occur in practice.

First, individual DPIs can stop being fingerprintable if they converge on the same, ``mainstream'' behaviors for parsing and interpreting traffic ambiguities. A parallel example can be found in TLS: certain censorship circumvention tools now try to mimic the clienthello structures (\eg, ciphersuites, extensions, \etc) of popular browsers, effectively converging their fingerprints to the much larger set of browser-generated TLS traffic. For DPIs, however, achieving such convergence appears significantly less feasible. Vendors tend to disagree on how to interpret corner cases left ambiguous by RFCs (which is what enabled our work in the first place), and these differences reflect genuine uncertainty or differing priorities (\eg, fail-open vs. fail-close). Furthermore, the ``correct'' interpretation of traffic for DPIs can sometimes depend on the behavior of the end-host that ultimately receives the traffic. For example, operating systems vary widely in their handling of overlapping IP fragments~\cite{ptacek1998insertionsurvey2}, so the ``correct'' DPI behavior---that aligns with the endhost---is inherently context-specific. As such, there is no strong incentive or mechanism for DPI vendors to adopt identical ``reference'' implementations.

Another scenario involves DPI vendors actively deploying defenses to resist fingerprinting. For example, previous proposals suggest placing ``traffic normalizers'' upstream of DPIs to resolve ambiguities before the traffic reaches the DPI~\cite{handley2001networksurvey1, shankar2003activesurvey17}. Yet, this merely shifts the target: now the normalizer itself becomes the new fingerprinting target. Alternatively, DPIs could introduce randomness into their behaviors, analogous to Chrome's recent adoption of randomized TLS extension ordering to avoid having one static TLS fingerprint~\cite{chrometlsfingerprint}. For example, DPIs might introduce probabilistic blocking when triggered, deliberately adding noise to their fingerprints. Yet, such randomness would directly undermine the reliability of traffic filtering, resulting in occasional failures to block content that should be censored. We believe such proposition---sacrificing reliability to resist fingerprinting---is unlikely to be appealing to vendors given the fundamental goals of DPI deployment.

\subsection{Limitations}
\label{sec:limitation}
A key limitation of our fingerprinting methodology relates to the inherent constraints of remote measurement. In some cases, there may be a middlebox along the network path that partially or completely alters our probing traffic before it reaches the DPI of interest. For example, if a router en route reassembles IP fragments and sends only the reassembled packets onward, our IP-fragmentation-based probes will be nullified. More challenging still is the presence of multiple DPIs along the network path. If two distinct DPIs enforce overlapping censorship policies, our measured fingerprint will reflect a compound behavior from the two DPIs combined, and it would be hard to isolate and fingerprint either one independently. While we attempt to detect these situations with our \analyzer (Table~\ref{tab:analyzer}), fundamentally we cannot guarantee that each measured fingerprint corresponds solely to a single DPI deployment. 

While remote fingerprinting provides large scale, a limitation is that we cannot measure DPIs that apply censorship policies asymmetrically---DPIs that apply filtering selectively depending on the direction of the traffic. One such example that we are aware of is Russia's TSPU system~\cite{xue2022tspusurvey19}, which examines and enforces censorship on outbound connections originate from Russian hosts but does not filter inbound connections. In such cases, probing from an external vantage point may not trigger censorship, rendering the DPIs effectively invisible to our fingerprinting. Future work could apply our methodology from in-network vantage points to capture these behaviors.

\subsection{Fingerprinting Other Targeted Interference}
While this work has focused on DPI devices, the underlying fingerprinting approach can naturally extend to other middleboxes that selectively interfere with traffic---such as targeted throttling or TLS man-in-the-middle. Our fingerprinting methodology only requires two essential conditions: 1) the middlebox inspects and interprets packet flows to evaluate the configured policy (\ie, indiscriminate interference, such as complete Internet shutdowns, fall outside the scope); and 2) once triggered, the interference is externally observable (\eg, RST/blockpage injection for censorship, lowered throughput for throttling, or altered TLS certificates for MITM). Future work can adapt this generalizable methodology to characterize and understand other types of interfering middleboxes.

\subsection{Ethics}

The measurement methodology of \nameofthething, which sends crafted probes containing potentially censored keywords to intentionally trigger DPI responses, requires careful ethical consideration. In line with community best practices, we consulted our institution's IRB, took measures to minimize the impact on the targeted networks, and clearly identified our measurement infrastructure for transparency. A more detailed discussion of the ethical considerations can be found in Appendix~\ref{sec:ethics}.

%% file: 07conclusion.tex
\section{Conclusion}
\label{sec:conclusion}

In this paper, we explore the practicality of exploiting the inherent ambiguities in traffic parsing and interpretation to derive behavioral fingerprints for DPI devices. Our experiments demonstrate that even a modest set of 20-40 carefully crafted packet sequences (``probes'') provides sufficient discriminative power to effectively differentiate and cluster black-box DPI implementations. We hope our work expands the community's visibility into these traffic-interfering middleboxes and encourages broader measurement efforts toward greater transparency and accountability of DPI deployments across the global Internet.

%% file: 99appendix.tex
\appendix

\section{Appendix}
\label{sec:appendix}

\subsection{Ethics}
\label{sec:ethics}

Our measurements involve sending crafted traffic (with potentially blocked domain name or keywords) to remote endpoints in order to trigger and observe DPI behaviors, which raises ethical considerations regarding potential harms such measurements may cause. In developing this work, we sought guidance from our institution's IRB, presenting our research plan and measurement methodology in detail. Consistent with prior measurement studies on similar subjects~\cite{raman2022cenfuzzsurvey7, GFWfullyencrypted, Fan2025a}, the IRB determined that our study does not involve human subjects and granted a ``Not Regulated'' determination. It is important to emphasize, however, that an IRB exemption is not an endorsement or approval of the study's ethics but merely indicates that the study does not meet the criteria for ``human subjects research'' and thus falls outside their oversight. Therefore, the responsibility lies with us as researchers to adopt measures to minimize risks and potential harms.

We followed community norms for large-scale Internet measurement~\cite{GFWfullyencrypted, xue2022tspusurvey19, raman2022cenfuzzsurvey7} by setting up a dedicated webpage and reverse DNS records on our measurement machines to identify ourselves, explain the purpose of our research, and provide contact information. We carefully ensure not to overwhelm endpoints selected as measurement targets by running measurements one probe at a time for each target---with probes averaging less than 1 kilobyte---and spacing probes 120 seconds apart to avoid straining the network resources of the targets.

We also note that half of our probes (non-Control ones) intentionally include a potentially blocked domain name or censored keyword to elicit DPI responses. While the potential risks involved in censorship measurements remain an area of debate within the measurement community~\cite{Jones2015a, Narayanan2015a}, we emphasize that all our measurements are fully remote, meaning that all connections are initiated exclusively from our measurement machines located on the ``external/public'' side of the DPI devices, with targeted web servers merely accepting inbound connection requests on their listening ports. We consider it highly unlikely that receiving unsolicited traffic with censored keywords could implicate these web servers, given that such servers, publicly accessible on the Internet, inherently have no control over the types of traffic they receive. 

\subsection{Open Science}
\label{sec:openscience}

The source code of \nameofthething can be found at \url{https://github.com/censoredplanet/CenDPI}.

\subsection{Reduce \textit{Inconclusive} Measurement Outcomes}
\label{sec:inconclusive}
To decrease the incidence of inconclusive measurements---such as the \{R2,R3,R4\}\{R1\} pattern in Table 2---we can append a standard application-layer request with the Control Domain after the mutated request packet. If the web server discards the mutated request packet but still responds to the appended standard request, we can more easily distinguish between server-side rejection and DPI-induced blocking. For example, suppose a particular mutation causes the target web server itself to drop the request. Without an appended standard request, the resulting blackhole matches the DPI’s triggered blocking behavior, producing the ambiguity discussed in \S~\ref{sec:analyzer}. However, if we append a standard control request to the same flow, then in R3 (Mutated Control) we might receive a legitimate server response to that appended request, whereas in R4 (Mutated Test) we would still see no response if the DPI was triggered. By comparing outcomes across R2, R3, and R4, \analyzer can more conclusively determine whether the DPI has been triggered, or if the server simply dropped malformed packets.

One caveat is that this approach assumes the DPI enforces session-level blocking. That is, once a DPI is triggered by a violating packet, it typically blocks the entire connection flow. If a stateless DPI drops only the offending packet but then allows subsequent packets, we might mistakenly label the probe as a bypass. While this represents a false negative in evasion attack discovery, it does not critically undermine our \textit{fingerprinting} goal, which aims for consistent signatures across identical DPI implementations. As long as we reliably produce the same verdict in these corner cases, the method remains valid for clustering and differentiation (\ie, a stateless DPI would still exhibit its own DPI fingerprint that is likely very different from any stateful DPI).

Finally, not all inconclusive patterns can be resolved by appending extra requests---in some cases, web servers may terminate connections immediately upon receiving a mutated packet (\eg, injecting a RST). However, at the probe selection stage, we can filter out probes that frequently yield such inconclusive outcomes and prioritize those that provide more conclusive verdicts.

\input{listings/example_probe}
\input{listings/tcpip}

\clearpage
\subsection{Root Cause Analysis}
Figures~\ref{fig:zeek63}, \ref{fig:snort63}, and \ref{fig:snort55} show excerpts from the Zeek and Snort source code that illustrate the specific behaviors we leverage for fingerprinting.

\input{figures/zeek_63}

\input{figures/snort_63}

\input{figures/snort_55}

\input{tables/probes40}

%% file: listings/example_probe.tex
\subsection{Example Probe Configuration}
\label{sec:exampleprobe}

\lstset{
    basicstyle=\small\ttfamily, 
    frame=single,
    breaklines=true,
    escapeinside={(*@}{@*)},
}

\begin{lstlisting}[language={},caption={A sample YAML probe configuration for \nameofthething, testing DPI behaviors when encountering packets with invalid (too old) TCP Timestamp option.},label={lst:yaml-probe}]
protocol: http/https
applicationMessage:
  http: 
    request: "GET / HTTP/1.1\r\nHost: ${}\r\nUser-Agent: curl/8.11.1\r\nAccept: */*\r\n\r\n"
  tls: 
    clientHelloConfig:
      chVersion: "0303"
    records:
      - contentType: "16"
        recordVersion: "0301"
        payloadType: "clienthello"
        offset: 0
        length: -1

packets:
  - ethernet:
    ip:
      tos: 0
      ttl: 64
      id: 33345
      protocol: tcp
      moreFragments: false
      dontFragment: true
      ipOptions:
    tcp:
      window: 65535
      urgentPointer: 0
      flags:
        syn: true
      tcpOptions:
        - tcpOptionType: 1
        - tcpOptionType: 1
        - tcpOptionType: 2
          tcpOptionLength: 4
          tcpOptionData: "05B4"
        - tcpOptionType: 3
          tcpOptionLength: 3
          tcpOptionData: "06"
        - tcpOptionType: 8
          tcpOptionLength: 10
          tcpOptionData: "0102030000000000"
    delay: 1
      
  - ethernet:
    ip:
      tos: 0
      ttl: 64
      id: 33346
      protocol: tcp
      moreFragments: false
      dontFragment: true
      ipOptions:
    tcp:
      window: 2056
      urgentPointer: 0
      flags:
        ack: true
      tcpOptions:
        - tcpOptionType: 8
          tcpOptionLength: 10
          tcpOptionData: "0102030400000000"

  - ethernet:
    ip:
      tos: 0
      ttl: 64
      id: 33347
      protocol: tcp
      moreFragments: false
      dontFragment: true
      ipOptions:
    tcp:
      window: 2056
      flags:
        psh: true
        ack: true
      messageOffset: 0
      messageLength: -1
      tcpOptions:
        - tcpOptionType: 8
          tcpOptionLength: 10
          tcpOptionData: "0102000000000000"
    delay: 1

  - ethernet:
    ip:
      tos: 0
      ttl: 64
      id: 33348
      protocol: tcp
      moreFragments: false
      dontFragment: true
      ipOptions:
    tcp:
      window: 2056
      flags:
        psh: true
        ack: true
      messageOffset: 0
      messageLength: -1
      reverseDomain: true
      tcpOptions:
        - tcpOptionType: 8
          tcpOptionLength: 10
          tcpOptionData: "0102030500000000"
    delay: 1

  - ethernet:
    ip:
      tos: 0
      ttl: 64
      id: 33349
    tcp:
      window: 2056
      flags:
        ack: true
        fin: true
    delay: 1

  - ethernet:   
    ip:
      tos: 0
      ttl: 64
      id: 33350
    tcp:
      window: 2056
      flags:
        ack: true
\end{lstlisting}

%% file: listings/tcpip.tex
%\vspace{90mm}
\pagebreak
\subsection{TCP\&IP Mutations}
\label{sec:tcpipmutations}

\lstset{
    %frame=single,
    basicstyle=\small\ttfamily, 
    breaklines=true,
    escapeinside={(*@}{@*)},
}
\begin{lstlisting}[language={},caption={Fields in TCP and IP that we mutate to generate candidate probes. Red: previously exploited in evasion attacks and are mutated in this work; blue: previously exploited fields that we do not mutate.},label={lst:tcpip}]
IP:
    Version
    Differentiated Service
    (*@\textcolor{myBlue}{Total Length} @*)
    (*@\textcolor{myRed}{Identification} @*)
    (*@\textcolor{myRed}{IP Flags (Reserved, DF, MF)} @*)
    (*@\textcolor{myRed}{Fragment Offset} @*)
    (*@\textcolor{myBlue}{TimeToLive} @*)
    (*@\textcolor{myRed}{Protocol} @*)
    (*@\textcolor{myRed}{Checksum} @*)
    Source Address
    Destination Address
    (*@\textcolor{myRed}{IP Options:} @*)
        (*@\textcolor{myRed}{IP Options Type} @*)
        (*@\textcolor{myBlue}{IP Options Length} @*)
        (*@\textcolor{myRed}{IP Options Value} @*)
    (*@\textcolor{myRed}{Payload} @*)
TCP:
    Source Port
    Destination Port
    (*@\textcolor{myRed}{Sequence Number} @*)
    (*@\textcolor{myRed}{Acknowledge Number} @*)
    (*@\textcolor{myBlue}{Length} @*)
    (*@\textcolor{myRed}{TCP Flags} @*)
    (*@\textcolor{myRed}{Window} @*)
    (*@\textcolor{myRed}{Checksum} @*)
    (*@\textcolor{myRed}{Urgent Pointer Value} @*)
    (*@\textcolor{myRed}{TCP Options:} @*)
        (*@\textcolor{myRed}{TCP Options Type} @*)
        (*@\textcolor{myBlue}{TCP Options Length} @*)
        (*@\textcolor{myRed}{TCP Options Value} @*)
    (*@\textcolor{myRed}{Payload} @*)
\end{lstlisting}

%% file: figures/zeek_63.tex
\begin{figure}[t]
\centering
 \includegraphics[width=\columnwidth,keepaspectratio]{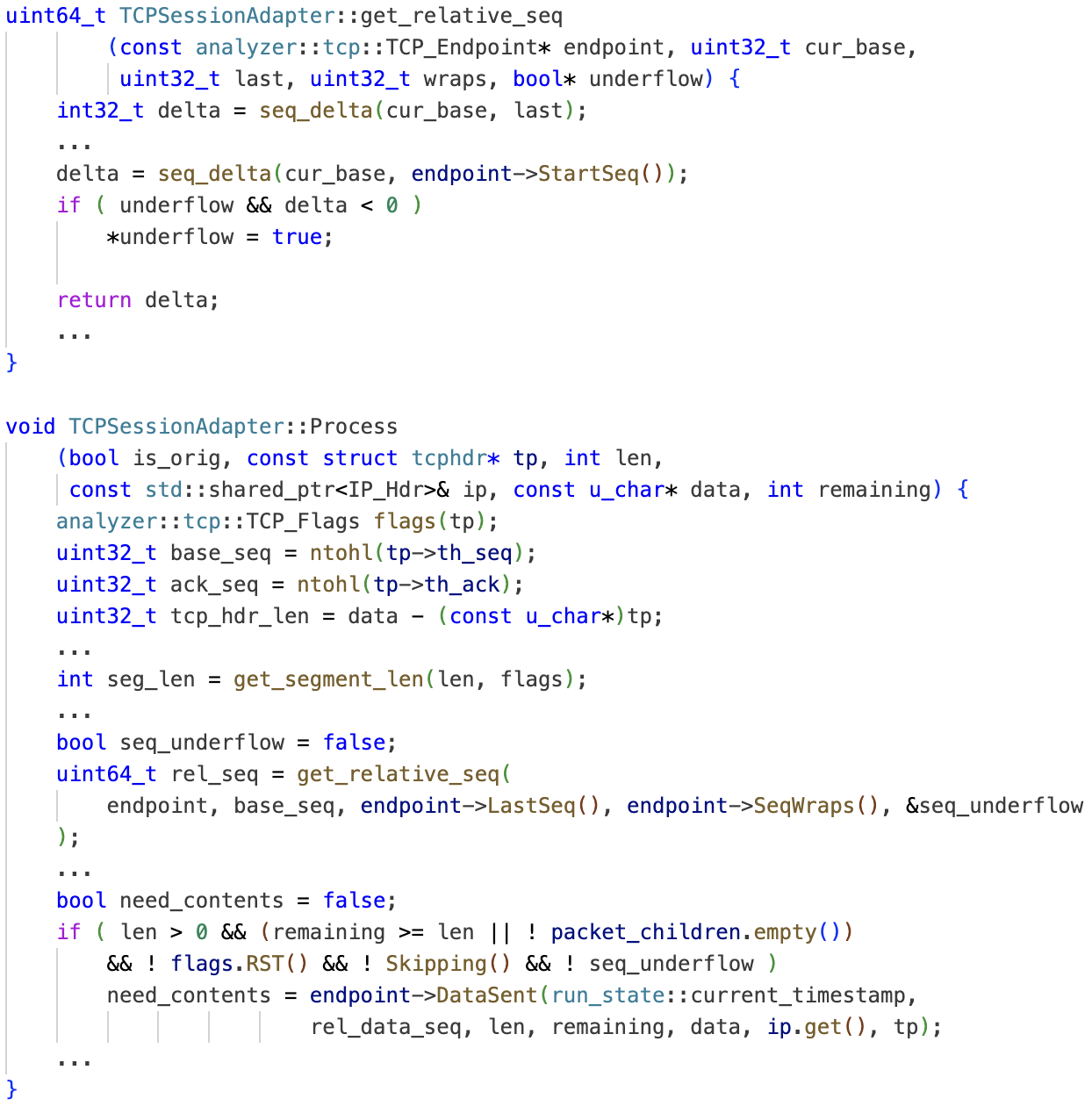}
\caption{Zeek's (v7.0.4) handling of partially out-window TCP segments.}
\label{fig:zeek63}
\vspace{-10pt}
\end{figure}

%% file: figures/snort_63.tex
\begin{figure}[t]
\centering
 \includegraphics[width=\columnwidth,keepaspectratio]{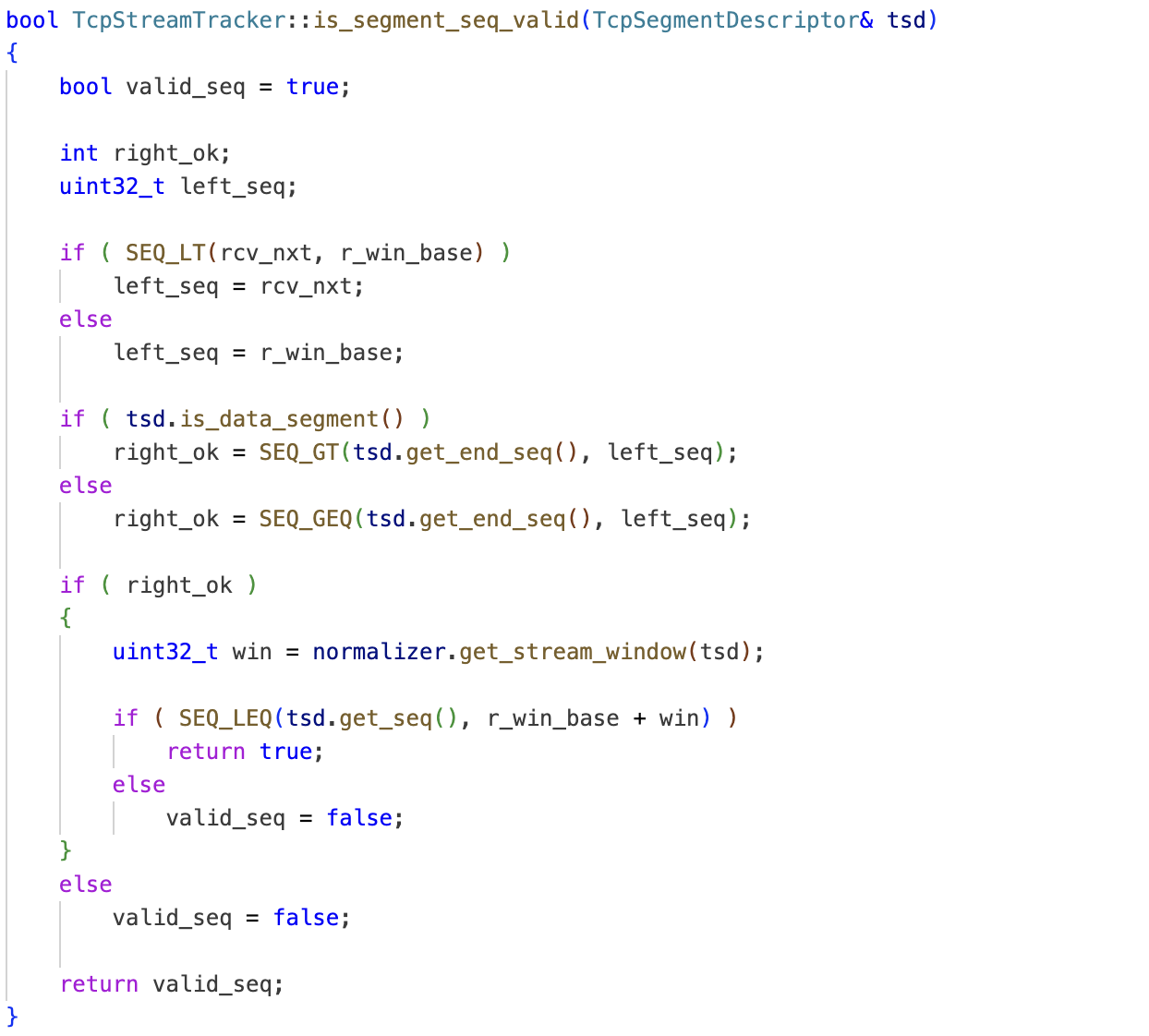}
\caption{Snort's (v3.6.0) handling of partially out-window TCP segments.}
\label{fig:snort63}
\vspace{-10pt}
\end{figure}

%% file: figures/snort_55.tex
\begin{figure}[t]
\centering
 \includegraphics[width=\columnwidth,keepaspectratio]{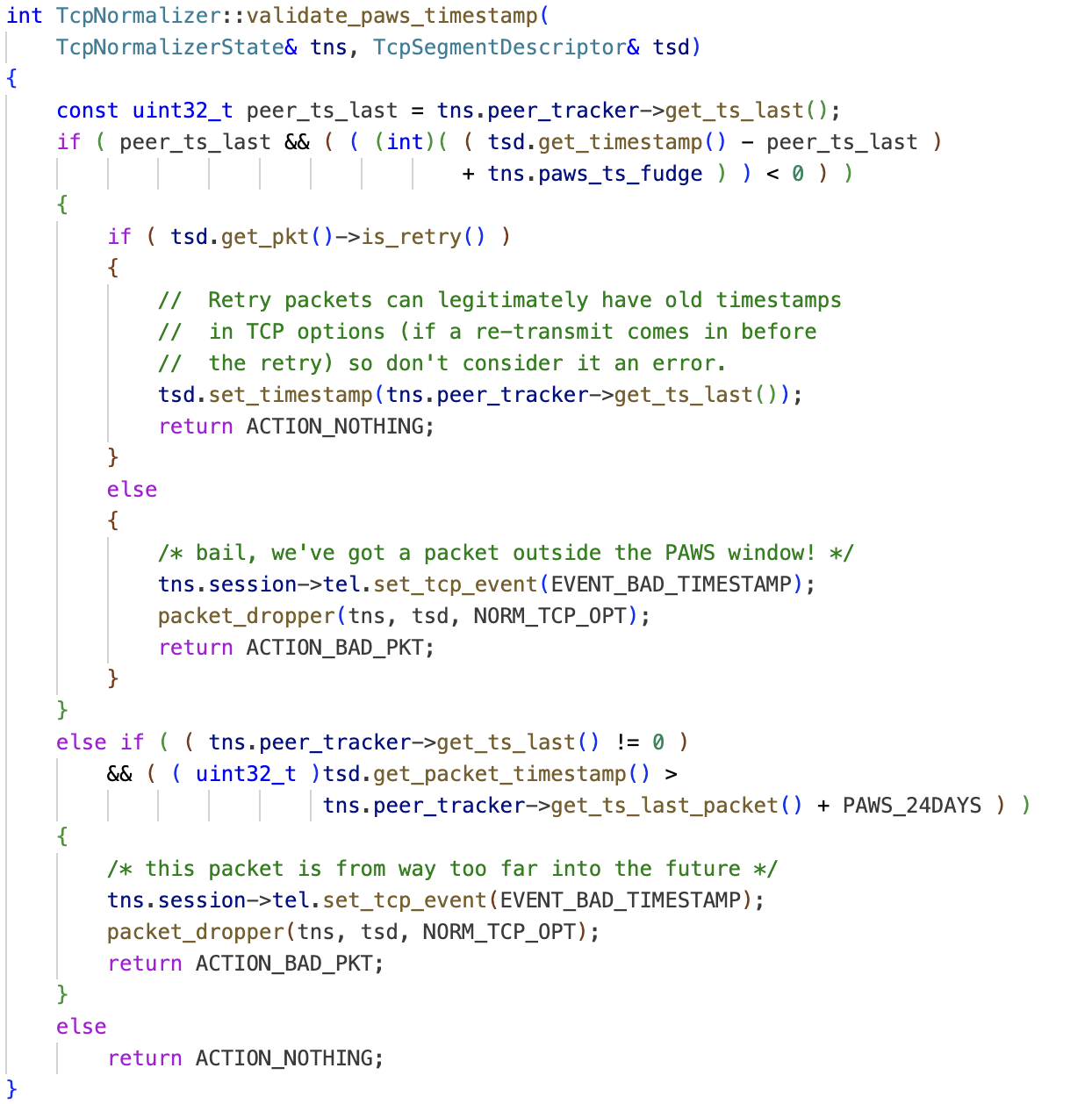}
\caption{Snort's (v3.6.0) TCP Timestamp validation.}
\label{fig:snort55}
\vspace{-10pt}
\end{figure}

%% file: tables/probes40.tex
\begin{table*}[t!]
\footnotesize
\centering

\begin{tabular*}{2\columnwidth}{@{\extracolsep{\fill}}m{0.5cm}m{1cm}m{5.4cm}m{9cm}}

\toprule
\multicolumn{4}{c}{\textbf{Common Probes}} \\
\toprule
        {Layer} & {Type} & {Name} & {Description} \\
\midrule

%31
IP & Fragment & Fragment[l:IP;t:maxDist;maxdist:16] &  Split the triggering request at the IP layer into two fragments; send the first fragment; then send 16 dummy fragments with random data with the same IP addresses but different IPID; finally send the second fragment of the triggering request. \\ \midrule

%141
TCP & Insert & Insert[p:I3;f:P;d:altProto;option:] & Insert a TCP packet containing a non-triggering request of the other protocol (\ie, send a HTTP GET if the current measurement is HTTPS, or a TLS clienthello if HTTP), before sending the triggering request.  \\ \midrule

%63
TCP & Mutate & Mutate[l:TCP;f:seq;option:negativeSeqWithPadding] & Mutate the triggering packet with SEQ $\leq$ ISN, and prepend the payload with padding so that the request data is in-window. \\ \midrule

%4
IP & Fragment & Fragment[l:IP;t:outorder] & Split the triggering request at the IP layer into three fragments and send them backwards. \\ \midrule

%22
IP & Fragment & Fragment[l:IP;t:overlapping;position:lshortrequal] & Split the triggering request at the IP layer into multiple fragments, with two fragments partially overlap. In this case, the second overlapping fragment has a larger offset and ends exactly at the right boundary of the first fragment. \\ \midrule

%40
IP & Mutate & Mutate[l:IP;f:option;option:noop] & Add a NO-Operation (NOP) option to the IP header of the triggering request.\\ \midrule

%44
TCP & Mutate & Mutate[l:TCP;f:checksum;checksum:corrupt] & Corrupt the TCP checksum of the triggering request. \\ \midrule

%60
TCP & Mutate & Mutate[l:TCP;f:urgentPointer;option:noack] & Add the Urgent TCP flag to the TCP header of the triggering request while removing the ACK flag. Use a random ($\leq len(payload)$) value as the value of the urgent pointer. \\ \midrule

%139
TCP & Insert & Insert[p:I3;f:P;d:controlRequest;option:] & Insert a TCP packet containing a non-triggering request, with only PSH flag (no ACK) before sending the triggering request. \\ \midrule

%55
TCP & Mutate & Mutate[l:TCP;f:option;option:timestamp] & Add a TCP Timestamp option to the TCP header of all outgoing packets. Then mutate the timestamp of the triggering request to a timestamp earlier than that of the preceding outgoing packet. \\ \midrule

%58
TCP & Mutate & Mutate[l:TCP;f:urgentPointer;option:] & Add the Urgent TCP flag to the TCP header of the triggering request. Use a random ($\leq len(payload)$) value as the value of the urgent pointer.\\ \midrule

%88
TCP & Fragment & Fragment[l:TCP;t:overlapping;position:lequalrlong] & Split the triggering request at the TCP layer into multiple segments, with two segments partially overlap. In this case, the second overlapping segment has the same offset as the first one but has a larger size so that it extends beyond the right boundary of the first segment.\\ \midrule

%138
TCP & Insert & Insert[p:I3;f:PA;d:controlRequest;option:checksum] & Insert a TCP packet containing a non-triggering request, with PSH/ACK flag and corrupt TCP checksum, before sending the triggering request.\\ \midrule

%12
IP & Fragment & Fragment[l:IP;t:fragmentNum;num:55] & Split the triggering request at the IP layer into 55 fragments.\\ \midrule

%37
IP & Mutate & Mutate[l:IP;f:flag;flags:M] & Change the IP flag of the triggering request to ``More Fragment''.\\ \midrule

%93
TCP & Fragment & Fragment[l:TCP;t:overlapping;position:llongrshort] & Split the triggering request at the TCP layer into multiple segments, with two segments partially overlap. In this case, the second overlapping segment has a smaller offset than the first one and ends within the right boundary of the first segment.\\ \midrule

%39
IP & Mutate & Mutate[l:IP;f:flag;flags:E] & Set the reserved bit (``evil bit'') in the IP header of the triggering request. \\ \midrule

%54
TCP & Mutate & Mutate[l:TCP;f:option;option:md5] & Add a MD5 signature (with invalid MD5 digest) option to the TCP header of the triggering request. \\ \midrule

%68
TCP & Fragment & Fragment[l:TCP;t:fragmentSize;size:8] & Split the triggering request at the TCP layer into two segments, with the first segment being 8-byte long. \\ \midrule

%27
IP & Fragment & Fragment[l:IP;t:overlapping;position:llongrlong] & Split the triggering request at the IP layer into multiple fragments, with two fragments partially overlap. In this case, the second overlapping fragment has a smaller offset and a larger end position so that it entirely ``wraps'' the first fragment inside it. \\ \midrule

%86
TCP & Fragment & Fragment[l:TCP;t:overlapping;position:lshortrequal] & Split the triggering request at the TCP layer into multiple segments, with two segments partially overlap. In this case, the second overlapping segment has a larger offset than the first one and ends exactly at the right boundary of the first segment. \\ \midrule

\toprule
\multicolumn{4}{c}{\textbf{HTTP-only Probes}} \\
\toprule

%23
IP & Fragment & Fragment[l:IP;t:overlapping;position:lshortrshort] & Split the triggering request at the IP layer into multiple fragments, with two fragments partially overlap. In this case, the second overlapping fragment has a larger offset and a smaller end position so that it entirely ``wrapped'' by the first fragment. \\ \midrule

%169
HTTP & Mutate & Mutate[l:App;t:http;f:version;value:HTTP: 1.1] & Mutate the version field of the triggering HTTP request. Use ``HTTP: 1.1'' as version (additional space in the version value). \\ \midrule

%94
TCP & Fragment & Fragment[l:TCP;t:overlapping;position:lequalrequal] & Split the triggering request at the TCP layer into multiple segments, with two segments completely overlap. \\ \midrule

%162
HTTP & Mutate & Mutate[l:App;t:http;f:method;value:GE] & Mutate the HTTP Method field of the triggering HTTP request. Use ``GE'' as the request method. \\ \midrule

%722
TCP & Insert & Insert[p:I1;f:A;d:controlRequest;option:md5] & Insert a TCP packet containing a non-triggering request, with only ACK flag (no PSH) and an invalid MD5 option before sending the triggering request. \\ \midrule

%52
TCP & Mutate & Mutate[l:TCP;f:flag;flags:SAFPU] & Set the TCP flags of the triggering request to (SYN, ACK, FIN, PSH, URG). \\ \midrule

\end{tabular*}

\caption{\textbf{Top 40 probes selected following the procedure described in \S~\ref{sec:probeselection}.}}
\label{tab:selectedprobes}
\vspace{-10pt}
\end{table*}

\begin{table*}[t!]
\footnotesize
\centering

\begin{tabular*}{2\columnwidth}{@{\extracolsep{\fill}}m{0.5cm}m{1cm}m{5.4cm}m{9cm}}

\toprule
\multicolumn{4}{c}{\textbf{HTTP-only Probes (continued)}} \\
\toprule

%145
HTTP & Mutate & Mutate[l:App;t:domain;c:prepend;char:star] & Prepend stars (*) before the domain name of the triggering request. \\ \midrule

%171
HTTP & Mutate & Mutate[l:App;t:http;f:version;value:HTTP:3] & Mutate the version field of the triggering HTTP request. Use ``HTTP:3'' as version. \\ \midrule

%155
HTTP & Mutate & Mutate[l:App;t:http;f:delimiter;char:09] & Replace the default delimiter (simple space) in the triggering request with horizontal tabs (x09). \\ \midrule

%1046
TCP & Insert & Insert[p:I1;f:PU;d:controlRequest;option:timestamp] & Insert a TCP packet containing a non-triggering request, with PSH and URG flag set and an invalid timestamp option after sending the initial SYN. \\ \midrule

%101
TCP & Insert & Insert[p:I3;f:R;d:;option:checksum] & Insert a TCP packet containing no payload, with RST flag set and an invalid checksum before sending the triggering request. \\ \midrule

%152
HTTP & Mutate & Mutate[l:App;t:http;f:request;option:tworequest] & Have two http requests in a single TCP packet; with the triggering request being the second one. \\ \midrule

%156
HTTP & Mutate & Mutate[l:App;t:http;f:delimiter;char:0b] & Replace the default delimiter (simple space) in the triggering request with vertical tabs (x0b). \\ \midrule

%163
HTTP & Mutate & Mutate[l:App;t:http;f:method;value:GeT] & Mutate the HTTP Method field of the triggering HTTP request. Use ``GeT'' as the request method. \\ \midrule

%1627
TCP & Insert & Insert[p:I3;f:R;d:controlRequest;option:checksum] & Insert a TCP packet containing a non-triggering request, with RST flag set and an invalid checksum before sending the triggering request. \\ \midrule

%153
HTTP & Mutate & Mutate[l:App;t:http;f:delimiter;char:r] & Replace the default line delimiter (\textbackslash r\textbackslash n) in the triggering request with only \textbackslash r. \\ \midrule

%154
HTTP & Mutate & Mutate[l:App;t:http;f:delimiter;char:n] & Replace the default line delimiter (\textbackslash r\textbackslash n) in the triggering request with only \textbackslash n. \\ \midrule

%1047
TCP & Insert & Insert[p:I1;f:PU;d:controlRequest;option:checksum] & Insert a TCP packet containing a non-triggering request, with PSH and URG flags set and an invalid checksum after sending the initial SYN. \\ \midrule

%1068
TCP & Insert & Insert[p:I1;f:PAU;d:controlRequest;option:] & Insert a TCP packet containing a non-triggering request, with PSH, ACK, and URG flags set before sending the initial SYN. \\ \midrule

\toprule
\multicolumn{4}{c}{\textbf{HTTPs-only Probes}} \\
\toprule

%132
TCP & Insert & Insert[p:I3;f:PA;d:random;option:checksum] & Insert a TCP packet containing random bytes as payload, with PSH and ACK flags set and corrupted checksum before sending the triggering request. \\ \midrule

%148
HTTPS & Mutate & Mutate[l:App;t:domain;c:append;char:space] & Append spaces to the SNI of the triggering Clienthello. \\ \midrule

%150
HTTPS & Mutate & Mutate[l:App;t:tls;f:recordVersion;value:0304] & Set the record-layer version of the triggering Clienthello to \textbackslash x03\textbackslash x04 \\ \midrule

%1042
TCP & Insert & Insert[p:I1;f:PU;d:controlRequest;option:outwindowSeq] & Insert a TCP packet with a non-triggering request, with PSH and URG flags set and an out-window sequence number after sending the initial SYN. \\ \midrule

%50
TCP & Mutate & Mutate[l:TCP;f:flag;flags:P] & Set the TCP flags of the triggering request to only PSH (no ACK). \\ \midrule

%81
TCP & Fragment & Fragment[l:TCP;t:fragmentNum;num:8] & Split the triggering request at the TCP layer into 8 segments.\\ \midrule

%83
TCP & Fragment & Fragment[l:TCP;t:fragmentNum;num:32] & Split the triggering request at the TCP layer into 32 segments.\\ \midrule

%89
TCP & Fragment & Fragment[l:TCP;t:overlapping;position:lequalrequal] & Split the triggering request at the TCP layer into multiple segments, with two segments completely overlap.\\ \midrule

%146
HTTPS & Mutate & Mutate[l:App;t:domain;c:append;char:star] & Append stars (*) to the SNI of the triggering Clienthello. \\ \midrule

%147
HTTPS & Mutate & Mutate[l:App;t:domain;c:prepend;char:space] & Prepend spaces to the SNI of the triggering Clienthello. \\ \midrule

%1891
TCP & Insert & Insert[p:I3;f:PU;d:controlRequest;option:inwindowSeq] & Insert a TCP packet with a non-triggering request, with PSH and URG flags set and an in-window but incorrect sequence number before sending the triggering request. \\ \midrule

%102
TCP & Insert & Insert[p:I3;f:RA;d:;option:checksum] & Insert a TCP packet with no payload, with RST and ACK flags set and a corrupted checksum before sending the triggering request. \\ \midrule

%151
HTTPS & Mutate & Mutate[l:App;t:tls;f:recordVersion;value:03ff] & Set the record-layer version of the triggering Clienthello to \textbackslash x03\textbackslash xff \\ \midrule

%153
HTTPS & Mutate & Mutate[l:App;t:tls;f:legacyVersion;value:0000] & Set the legacy version within the triggering Clienthello to \textbackslash x00\textbackslash x00 \\ \midrule

%1480
TCP & Insert & Insert[p:I2;f:PU;d:controlRequest;option:outwindowSeq] & Insert a TCP packet with a non-triggering request, with PSH and URG flags set and an out-window sequence number before sending the ACK that concludes the TCP handshake. \\ \midrule

%1454
TCP & Insert & Insert[p:I2;f:RP;d:controlRequest;option:checksum] & Insert a TCP packet with a non-triggering request, with RST and PSH flags set and corrupted checksum before sending the ACK that concludes the TCP handshake. \\ \midrule

%1478
TCP & Insert & Insert[p:I2;f:PU;d:controlRequest;option:checksum] & Insert a TCP packet with a non-triggering request, with PSH and URG flags set and corrupted checksum before sending the ACK that concludes the TCP handshake. \\ \midrule

%1679
TCP & Insert & Insert[p:I3;f:P;d:altProto;option:outwindowSeq] & Insert a TCP packet containing a non-triggering request of the other protocol (\ie, send a HTTP GET if the current measurement is HTTPS, or a TLS clienthello if HTTP) with only PSH flag set and an out-window sequence number, before sending the triggering request. \\ \midrule

%49
TCP & Mutate & Mutate[l:TCP;f:flag;flags:] & Remove all TCP flags of the triggering request. \\ \midrule

\bottomrule

\end{tabular*}

\caption{\textbf{(Continued) Top 40 probes selected following the procedure described in \S~\ref{sec:probeselection}.}}
\vspace{-10pt}
\end{table*}